\documentclass[aps,prd,nofootinbib,floatfix,showpacs]{revtex4}
\usepackage{colordvi}
\usepackage{graphicx}
\usepackage{bm}
\usepackage{multirow}
\usepackage{enumerate}
\reversemarginpar
\def\<{\langle}
\def\>{\rangle}
\newcommand{\text}{\rm}

\def\tr{{\text tr}\,}
\def\Eq#1{Eq.~(\ref{#1})}

\def\tilde{\widetilde}

\newcommand\pv{{\displaystyle{-} \hspace{-0.4cm} \int}}

\def\tr{{\text tr}\,}

\def\Eq#1{Eq.~(\ref{#1})}

\begin{document}

\vspace*{0.7in}
 
\begin{center}
{\large\bf Volume dependence of two-dimensional large-$N$ QCD\\
with a nonzero density of baryons.
}

\vspace*{1.0in}
\vspace{-1cm}
{Barak Bringoltz\\
\vspace*{.2in}
Department of Physics, University of Washington, Seattle,
WA 98195-1560, USA\\
}
\end{center}

\begin{abstract}

We take a first step towards the solution of QCD in $1+1$ dimensions at nonzero density. We regularize the theory in the UV by using a lattice and in the IR by putting the theory in a box of spatial size $L$. After fixing to axial gauge we use the coherent states approach to obtain the large-$N$ classical Hamiltonian ${\cal H}$ that describes color neutral quark-antiquark pairs interacting with spatial Polyakov loops in the background of baryons. Minimizing ${\cal H}$ we get a regularized form of the `t Hooft equation that depends on the expectation values of the Polyakov loops. Analyzing the $L$-dependence of this equation we show how volume independence, \`a la Eguchi and Kawai, emerges in the large-$N$ limit, and how it depends on the expectation values of the Polyakov loops. We describe how this independence relies on the realization of translation symmetry, in particular when the ground state contains a baryon crystal. Finally, we remark on the implications of our results on studying baryon density in large-$N$ QCD within single-site lattice theories, and on some general lessons concerning the way four-dimensional large-$N$ QCD behaves in the presence of baryons.

\end{abstract}


\maketitle

\setcounter{page}{1}
\newpage
\pagestyle{plain}

\section{Introduction}

QCD simplifies in the 't~Hooft limit of a large number of colors, and
as a result it has been a long-standing goal to understand the
properties of the theory in that limit~\cite{largeN}, including on the
lattice~\cite{lattice-reviews}.  One alternative to conventional large volume simulations is the use
the large-$N$ equivalence of QCD at large volume to QCD with zero volume
\cite{EK,BHN1,MK,Migdal,TEK,AEK,DW,GK,Parisipapers} (see also the related Ref.~\cite{KNN}).
These large-$N$ volume reductions allows one, in principle, to study very large values of $N\sim
O(100-400)$ with modest resources. 

Volume reduction holds only if the ground states of the large and zero volume theories respect certain symmetries~\cite{AEK}. Unfortunately, in the most interesting case of QCD in four dimensions these symmetries spontaneously break in the continuum limit  when a naive reduction prescription is used \cite{BHN1,MK}. An extension of that prescription is thus required and for a recent summary of the literature on this topic we refer to the reviews in the recent Refs.~\cite{QEK,DEK}.

In the case of two space-time dimensions -- the `t Hooft model -- a naive large-$N$ volume reduction is expected to hold and so this theory is generally thought to be completely independent of its volume. In the current paper we analyze this volume dependence. Our motivation is two-fold. Firstly, the `t Hooft model is analytically soluble at large-$N$. Thus we can explicitly see how large-$N$ volume reduction works in this case, and what may cause it to fail. This topic was also addressed for zero baryon number by the authors of Ref.~\cite{SchonThies_decompact}, and our treatment here differs from that paper by being manifestly gauge invariant, by going beyond zero baryon number, and by using the lattice regularization. Our approach also makes a direct connection with Eguchi-Kawai reduction, and shows how the expectation values of spatial Polyakov loops play a crucial role in the validity of volume independence.

Secondly, this paper is a prelude to our companion publication Ref.~\cite{nonzeroBpaper} where we use the formalism presented here to solve the theory in the presence of nonzero baryon density. Considering the current incomplete understanding of the way four-dimensional QCD behaves at low temperatures and large (but not asymptotic) baryon densities, we believe that such a study is useful. Also, there exist certain confusions in the literature about the way large-$N$ gauge theories behave at nonzero baryon number \cite{Cohen}, and seeing how these confusions go away in the soluble two-dimensional case is very helpful. 

Surprisingly, QCD in two dimensions and nonzero density has not been solved yet : While Ref.~\cite{Salcedo} studied only one and two baryons in an infinite volume, then Ref.~\cite{SchonThies} attempted to extend this but restricted to either (1) translational invariant states which were seen to be inconsistent, or (2) a particular translational non-invariant ansatz for a baryon crystal in the vicinity of the chiral limit. Since $1+1$-dimensional baryons become massless for massless quarks, it is natural to expect that they behave very differently than four-dimensional massive baryons. 
Furthermore, most of the current literature on the `t Hooft model has so far focused on its infinite volume limit where a certain set of gluonic zero modes are irrelevant. With a finite density of baryons, however, these become important and cannot be neglected (at least if the density is increased by fixing the baryon number and decreasing the volume). Thus, given the current status surveyed above, it seems wise to study the dense `t Hooft model for arbitrary quark mass, by making as few assumptions on the form of the ground state as possible, and by incorporating correctly the gluonic zero modes. In this paper we develop the machinery to achieve this goal. For the actual solution of the theory for arbitrary baryon numbers we refer to Ref.~\cite{nonzeroBpaper} and for all other discussions on the way nonzero chemical potential affects the system to Ref.~\cite{nonzeroMUpaper}.

Former studies of the `t Hooft model used a plethora of mathematical methods -- for example see \cite{plethora} for some papers relevant to this work. Common to all these is the need to control the severe IR divergences of this two-dimensional model. A particular clear approach, that we will follow in our study, is the one advocated in the seminal Refs.~\cite{LTYL,LNT}. There, one works in the Hamiltonian formalism defined in a spatial box of side $L$, and uses the axial gauge to remove all redundant degrees of freedom. This approach is also most suitable for our purpose of investigating the $L$-dependence of this Hamiltonian's ground state.

The outline of the paper is as follows. In Section~\ref{LQCDH} we present the details of the Hamiltonian approach to lattice QCD. A reader who is familiar with this approach can skip to Section~\ref{Haxial} where, by generalizing Refs.~\cite{LTYL,LNT} to the lattice, we show how to fix the axial gauge in the Hamiltonian formalism. Since such gauge fixing is less familiar than the gauge fixing in the Euclidean formalism, we do so in detail. Next, in Section~\ref{GLresolve}, we show how to resolve Gauss law and re-write the electric fields in terms of the fermion color charge densities. This rewriting can be done for all components of the electric field except for those conjugate to the eigenvalues of the spatial Polyakov loops. This set of eigenvalues and their conjugate electric fields is what we refer to above as zero modes, and in Section~\ref{0mode} we focus on them. Specifically, we show how to represent the zero modes' electric fields in the Schr\"oedinger picture as differential operators. 
The end result of Sections~\ref{LQCDH}-\ref{Hrecap} is a Hamiltonian that depends only on the fermions and on the zero-modes, with an overall color neutrality enforced on its Hilbert space. For the convenience of the reader we summarize this emerging structure in Section~\ref{Hrecap}. 
We then turn to find the ground state of this Hamiltonian. At large-$N$ this is done in two steps : (1) Solution of the gluon zero modes dynamics -  discussed in Section~\ref{SectorG}. (2) Solution of the fermion sector - Section~\ref{SectorF}. In the second step we use the coherent state approach of Refs.~\cite{YaffeCoherent} which seems particularly suitable for our problem. The end product is a regularized form of the `t Hooft classical Hamiltonian describing color neutral operators that correspond to quark-antiquark pairs and Polyakov-loops (that wrap around the spatial circle), and that interact in the presence of a fixed overall baryon number.\footnote{We note here that, in contrast to what happens in the lattice strong-coupling expansion \cite{MEandBEN}, the baryons are not tied to lattice sites in our approach and their wave-functions are determined dynamically.} In Section~\ref{otherworks} we survey other relevant works in the literature that obtain a similar Hamiltonian, pointing out the way they differ from our approach. In Section~\ref{decompact} we analyze the resulting Hamiltonian and its $L$ dependence for arbitrary baryon number $B$. We show how large-$N$ volume dependence emerges and that for it to hold we need to assume that the ground state has some degree of translation invariance. We also show how it can be violated by giving the Polyakov loops nonzero expectation values.\footnote{As discussed below, this will give rise to extra zero modes (in addition to the ones already mentioned above) but because the real ground state has vanishing Polyakov loops and no extra zero modes, we do not investigate this issue further.} An interesting phenomena occurs when the ground state contains a baryon crystal and we show how a `soft' form of volume independence emerges. This leads us to remark on the way our results affect studies of nonzero chemical potential that try to rely on large-$N$ volume independence. We conclude in Section~\ref{summary} by noting some general lessons one can learn about the way large-$N$ QCD behaves in the presence of baryons.

\section{Hamiltonian QCD in $1+1$ dimensions : a brief reminder }
\label{LQCDH}

In this section we introduce the Hamiltonian formalism of lattice QCD restricted to one spatial dimension and one flavor. A reader familiar with this formalism can skip to Section~\ref{Haxial}. 

This Hamiltonian of lattice QCD was first introduced by Kogut and Susskind in 1975 \cite{Kogut75}, shortly after Wilson's Euclidean formulation. This canonical formalism defines the theory of strong interactions on a spatial lattice with lattice spacing $a$ and continuous time $t$. In one dimension a lattice site is denoted by a single index $x$ taking integer values. We also use $x$ to denote the lattice link that is to the right of the site $x$. Because we define the theory on a finite box we set $x=1,2,\dots,L_s$, where $L_s=L/a$ is the number of lattice sites, and $L$ the physical length of the box. The boundary conditions of the gauge fields are taken to be periodic. In this section and throughout the paper we use standard lattice notations and so the factors of the lattice spacings are implicit and all fields are dimensionless.

The quantum fields that describe quarks are the fermion fields $\psi^{a\alpha}_{x}$ that live on the sites $x$ of the lattice. They have color indices $a=1,\dots,N$, and Dirac indices $\alpha=1,2$ (again, recall we are in $1+1$ dimensions). The Fermi fields obey the following anti-commutation relations
\begin{equation}
\left\{ \psi^{a\alpha}_x,\psi^{\dag b \beta}_y \right\} = \delta_{xy}\delta_{ab}\delta_{\alpha\beta}. \label{anticomm}
\end{equation}

Choosing to work in the temporal gauge that fixes $A_0=0$ removes one degree of freedom (and its conjugate momentum). This leaves the spatial gauge field operators $\left(U_x\right)_{ab}$ living on the lattice link between $x$ and $x+1$. For definiteness we note that the operator $\left(U^\dag_{x}\right)_{ab}$ is {\em defined} to be equal to $\left(\left(U_{x}\right)_{ba}\right)^\dag$, and so the following holds as a operator identity
\begin{equation}
\sum_a\, \left(\left(U_{x}\right)_{ab}\right)^\dag\, \left(U_{x}\right)_{ac} = \sum_a\, \left(U^\dag_{x}\right)_{ba}\, \left(U_{x}\right)_{ac}=\delta_{bc}.  
\end{equation}  
The conjugate momenta of the $U$'s also reside on links and are denoted by $E^i_{x}$, $i=1,\dots,N^2-1$. The following are the commutation relations of this set of operators,
\begin{eqnarray}
\left[ E^i_{x} , \left( U_{x} \right)_{ab} \right] &=& \left( \lambda^i U_{x} \right)_{ab}, \label{comm_plus1} \\
\left[ E^i_{x} , E^j_{x} \right] &=& i f^{ijk} E^k_{x}. \label{comm_plus2}
\end{eqnarray}
Here $\lambda^i$ are matrices that represent the traceless generators of $SU(N)$ in the fundamental representation, and $f^{ijk}$ are the structure constants of $SU(N)$. We choose a normalization where
\begin{eqnarray}
\tr \left(\lambda^i\lambda^j\right) &=& \frac12 \delta^{ij},\label{norm1}\\
\sum_{i=1}^{N^2-1} \, \lambda^i_{ab} \, \lambda^i_{cd} &=& \frac12 \left(\delta_{ad}\, \delta_{cb} - \frac1{N} \, \delta_{ab}\, \delta_{cd}\right).\label{norm2}
\end{eqnarray}

Let us now discuss the lattice Hilbert space. A general state $|\Omega \rangle$ is the direct product on all lattice sites,
\begin{equation}
|\Omega \rangle = | \Omega \rangle_G \otimes | \Omega \rangle_{\Psi}.
\end{equation}
Here the first factor is the projection of the state $|\Omega\>$ to the gauge field sector, while the second factor describes the fermionic sector. Any state $|\Omega\>_{\Psi}$ is the following direct product
\begin{equation}
|\Omega\>_{\Psi} = \prod_{\otimes x} \left( |\Omega\>_{\Psi} \right)_x.
\end{equation}
Concentrate on the Hilbert space of each site : the ``lowest'' state is the no-quantum ``drained'' state $|\textrm{dr}\rangle_x$, defined as
\begin{equation}
\psi^{a\alpha}_x|\textrm{dr}\rangle_x = 0. \label{drained}
\end{equation}
Applications of various $\psi^{\dag a\alpha}_x$ create the corresponding quarks on that site, 
\begin{equation}
|a\alpha \rangle_{x\Psi} = \psi^{\dag a\alpha}_x |\textrm{dr}\rangle_x.
\end{equation}
For the free theory, $\alpha=1$ correspond to creation of positive energy excitations and $\alpha=2$ corresponds to creation of negative energy excitations. To use the usual quark--anti-quark language we write for each site and color
\begin{equation}
\psi=\left( \begin{array}{c} b \\ d^{\dag} \end{array} \right).
\end{equation}
$b^{\dag}$ creates a quark and $d^{\dag}$ an anti-quark. From \Eq{drained} we see that $b$ and $d^{\dag}$ annihilate the drained state, which means that this state is empty of quarks, and filled with anti-quarks. The local baryon density operator is
\begin{equation}
B_x \equiv \frac1{N}\sum_{a=1}^{N} \left[ b^{\dag a} b^{a} - d^{\dag a} d^{a} \right]_x =\frac1{N} \sum_a\psi^{\dag\,a}_x \psi^a_x - 1. \label{B}
\end{equation}
According to \Eq{B}, the baryon number of the drained state is $-1$, corresponding to filling the site with anti-quarks. The vacuum $|0\rangle$ is the state with no quarks and no anti-quarks. This state is the filled Dirac sea on a single site and obeys,
\begin{equation}
b|0\rangle=d|0\rangle=0.
\end{equation}
The baryon number of this state is $B=0$. Because of the Pauli exclusion principle we cannot put too many fermions on a single site. For a single flavor theory the maximum number of local baryon number will be $1$, and is found only in the state $|\textrm{filled} \rangle$
\begin{equation}
b^{\dag}|\textrm{filled}\rangle=d|\textrm{filled}\rangle=0.
\end{equation}
Below we will see that gauge invariance puts more restrictions on the single site Hilbert space in order that it be color neutral.

Moving to the gauge Hilbert space, we also write it as a direct product of the form
\begin{equation}
|\Omega\>_G = \prod_{\otimes x} \left( |\Omega\>_G \right)_{x}.
\end{equation}
Next, we denote the state with no electric field $E$ by $|0\>_G$,
\begin{equation}
 E^i| 0 \>_G = 0, \qquad \forall i.
\end{equation}
Any application of the link operators $\left( U_{x} \right)_{ab}$ on $|0\>_G$, creates states which correspond to flux lines on the link $x$. The state $|0 \>_G$ is the only state with no flux at all. Using the fact that the electric field operators generate a $SU(N)$ algebra, one can distinguish between the different quantum states created by the link operators as follows. Define the quadratic Casimir operator
\begin{equation}
\vec{E}^2_{x}\equiv\sum_{i=1}^{N^2-1}E^{i2}_{x}. \label{E2}
\end{equation}
It is clear that the flux-less state $| 0 \>_G$ is an eigenstate of this Casimir, with zero eigenvalue.
Next, the commutations of the Casimir with the link operators are verified from \Eq{comm_plus1} to be
\begin{equation}
\left[ \vec{E}^2_{x},\left( U_{x} \right)_{ab} \right] = C_F\left( U_{x} \right)_{ab},
\end{equation}
where $C_F=(N^2-1)/2N$ is the Casimir operator in the fundamental representation.
This means that the state $\left[ \left( U_{x} \right)_{ab}| 0 \>_G \right]$ is also an eigenstate of~(\ref{E2}), with eigenvalue equal to $C_F$. One can now classify the states in $|\Omega \>_G$ according to their $\vec{E}^2$ eigenvalue. The result is a Hilbert space with a ladder-like structure. The lowest state is $|0\>_G$, with zero flux, and is a singlet of $SU(N)$. Repeated applications of the gauge field operators $\left( U_{x} \right)_{ab}$ create states with higher and higher values of flux and the operators $\vec{E}^2_{x}$ measures the flux on the link $x$. Indeed we shall see shortly that it is proportional to the (kinetic) energy of the gauge fields.

To complete the picture we now discuss gauge invariance. First recall that the starting point of this formalism was to choose the time-like gauge. This leaves only time-independent gauge transformations as a symmetry. The fermion operators belong to the fundamental representation of the gauge symmetry and transform as
\begin{equation}
\psi^a_x \rightarrow \left( V_x \right)^{ab} \psi^b_x, \label{gauge_F1}
\end{equation}
with $V\in SU(N)$ given in general by
\begin{equation}
V_x=\exp \left[ i\sum_{i=1}^{N^2-1} \theta^i_x \lambda^i\right].
\end{equation} 
Using the anti-commutation relations~(\ref{anticomm}), one can show that the quantum operator ${\cal V}_F$ that realizes \Eq{gauge_F1} in Hilbert space as
\begin{equation}
\psi^a_x \rightarrow {\cal V}_F \, \psi^a_x \, {\cal V}^\dag_F,
\end{equation}
is 
\begin{equation}
{\cal V}_F = \exp \left[ -i\sum_x \sum_{i=1}^{N^2-1} \theta^i_x \left( \psi^{\dag a}_x \lambda^i_{ab} \psi^b_x \right) \right]. \label{gauge_F2}
\end{equation}
The gauge fields transform according to 
\begin{equation}
\left( U_{x} \right)_{ab} \rightarrow {\cal V}_G\, \left(U_x\right)_{ab}\, {\cal V}^\dag_G = \left( V_x U_{x}  V^\dag_{x+1} \right)_{ab}.
\end{equation}
Using the commutation relations in \Eq{comm_plus1}, one shows that the quantum operator ${\cal V}_G$ that generates these rotations is given by\footnote{Here we take $\theta_x$ to be periodic in $x$.}
\begin{equation}
{\cal V}_G = \exp \left[ +i\sum_x \sum_{i=1}^{N^2-1} \theta^i_x  \left( E^i_{x} - \left(U^{\rm Adj.}_{x-1}\right)_{ji} \, E^j_{x-1}\right)\right], \label{gauge_G}
\end{equation}
where here the matrix of operators $\left(U^{\rm Adj.}_{x-1}\right)_{ji}$ is the link matrix in the adjoint representation, i.e.\footnote{Some useful properties of the operator $\left(U^{\rm Adj.}\right)_{ij}$ are that its representation in Hilbert space is Hermitian, $\left(U^{\rm Adj.}\right)_{ij}=\left(\left(U^{\rm Adj.}\right)_{ij}\right)^\dag$, and that $\sum_{k} \left(U^{\rm Adj.}\right)_{ik} \left(U^{\rm Adj.}\right)_{jk}=\delta_{ij}$. The latter relation means that for the operator identity $\left(U^{\rm Adj.}\, U^{\dag \rm Adj.} \right)_{ij}=\delta_{ij}$ to hold we need to define $\left(U^{\dag \rm Adj.}\right)_{ij}\equiv \left(U^{\rm Adj.}\right)_{ji}$.}
\begin{equation}
\left(U^{\rm Adj.}_{x}\right)_{ij}= 2 \, \tr \left( \lambda^i \, U_{x} \, \lambda^j \, U^\dag_{x}\right). 
\end{equation}

Putting \Eq{gauge_F2}, and \Eq{gauge_G} together, we see that the operator that induces gauge transformations is 
\begin{equation}
{\cal V}=\exp \left[ i\sum_x \sum_{i=1}^{N^2-1} \theta^i_x  \left( \rho^i_G - \rho^i_F \right)_x\right],\label{gauge_trnsf}
\end{equation}
where $\rho^i_{Fx}$, and $\rho^i_{Gx}$ are the color charge densities of the fermions and of the gauge fields. These two quantities are given by
\begin{eqnarray}
\rho^i_{Fx}&=&\psi^{\dag a}_x \lambda^i_{ab} \psi^b_x, \\
\rho^i_{Gx}&=&  E^i_x-\left( U^{\text{Adj.}}_{x-1}\right)_{ji} E^j_{x-1} \equiv D E^i_x, \label{rhoG} 
\end{eqnarray}

Since the lattice Hamiltonian is gauge invariant, we know that the generators of the gauge transformations commute with the Hamiltonian
\begin{equation}
\left[ H , \rho^i_{Gx} - \rho^i_{Fx} \right]=0, \hskip 1cm \forall i \quad \forall x.
\end{equation}
This means that we can choose to work with a basis that block diagonalizes $\rho^i_G-\rho^i_F$. This breaks the Hilbert space to separate sectors classified by their eigenvalue $\rho^i_{\rm external,x}\equiv \rho^i_G-\rho^i_F$ on the lattice. Each set of values $\rho^i_{{\text external},x}$ describes a different physical case, with a different {\em external} distribution of color charge (that can correspond, for example, to infinitely heavy quarks etc.). To describe the physics of zero external gauge fields, we work with the choice $\rho^i_{\text{external}}=0$.
Working in this subspace means that {\em all} physical states must be color singlets, since all gauge transformations are trivial. This means that in this sector the following equations hold as operators identities
\begin{equation}
 E^i_x-\left( U^{\text{Adj.}}_{x-1}\right)_{ji} E^j_{x-1} = \rho^i_{Fx}. \label{Glaw}
\end{equation}

Finally we write the lattice Hamiltonian $H$ of the $1+1$-dimensional  $SU(N)$ gauge theory with one flavor of fermions. $H$ is given by (in one dimension there is no magnetic field and so the plaquette term is identically zero)
\begin{equation}
H=H_E+H_F. \label{eq:H_initial}
\end{equation}
Here $H_E$ is the electric term, a sum over links $x$
of the $SU(N)$ Casimir operator
\begin{equation}
H_G=\frac{g^2}{2}\sum_{x=1}^{L_s}\left(E^i_{x}\right)^2.
\end{equation}
Next is the  fermion Hamiltonian,
\begin{equation}
H_F=-\frac{i}{2}\sum_{x=1}^{L_s}\psi^{\dag a\alpha}_x\, \left(\sigma_3\right)_{\alpha\beta} \, U_{x,ab} \, \psi^{b\beta}_{x+1}+h.c. + m\sum_x \psi^{\dag a\alpha}_x \,\left(\sigma_1\right)_{\alpha\beta} \, \psi^{a\beta}_x.  \label{eq:H_F}
\end{equation}
Here we choose a particular representation of the one-dimensional Dirac matrices using the Pauli matrices $\sigma_{1,3}$ and periodic boundary conditions on the fermions, i.e. $\psi_x=\psi_{x+L_s}$, and $\psi^\dag_x=\psi^\dag_{x+L_s}$.

Superficially, the first term in the Hamiltonian is symmetric under a $U(1)_R\times U(1)_L$ chiral symmetry which is explicitly broken by the mass term to the vector $U(1)$. One can, however, spin diagonalize the fermions by writing $\psi^{\alpha}_x\to \left(e^{-i\pi\, \sigma_2/2 } \sigma_3^x\right)_{\alpha\beta} \psi^{\beta}_x$ and see that 
\begin{equation}
H_F=-\frac{i}2\sum_{x}\psi^{\dag a\alpha}_x  \, U_{x,ab} \, \psi^{b\alpha}_{x+1}+h.c. + m\sum_x \, (-1)^x\, \psi^{\dag a\alpha}_x \, \left(\sigma_3\right)_{\alpha\beta} \, \psi^{a\beta}_x.  \label{eq:H_F_stag}
\end{equation}
Here, while the first term is invariant under a $U(2)$ group, the mass term is invariant only under a $U(1)\times U(1)$. In our actual calculations we will drop the second component of this spin-diagonalized Hamiltonian (the $\alpha=2$), and thus work with staggered fermions~\cite{Susskind}. To go back to Dirac fermions is easy and for our purpose it is useful to note that the continuum chiral condensate and baryon number local densities, at position $X$, $\left(\bar \psi\psi\right)(X)$ and $\left(\psi^\dag\psi\right)(X)$, are given, up to overall renormalization factors, by \cite{Susskind}
\begin{eqnarray}
\left(\psi^\dag \psi\right)(X)&=&\frac1{\sqrt{2}}\left(\psi^\dag_x \psi_x + \psi^\dag_{x-1} \psi_{x-1} \right),\label{psibarpsi}\\
\left(\bar \psi \psi\right)(X)&=&\frac1{\sqrt{2}}\left(\psi^\dag_x \psi_x - \psi^\dag_{x-1} \psi_{x-1} \right),\label{psidagpsi}
\end{eqnarray}
where we take $X/a=x$, and $x$ denotes an even site.\footnote{Here all fields are still dimensionless. The continuum expectation values $\frac{\<\bar \psi \left({\bm 1},\gamma_0\right)\psi\>^{\rm continuum}}{\sqrt{g^2N}}$ are given by dividing the r.h.s. of Eqs.~(\ref{psibarpsi})--(\ref{psidagpsi}) by $a\sqrt{g^2N}$ ($g$ has dimensions of mass in $1+1$).}

\section{Axial gauge fixing}
\label{Haxial}

In this section we show how to fix the axial gauge. 
Since axial gauge fixing in the Hamiltonian approach is less familiar than it is in the Euclidean approach we begin with an explanation of the general strategy.

 Temporal gauge fixing left us with a Hamiltonian that is invariant under time-independent gauge transformations discussed in the previous section. With no external charges the generator of such transforms needs to vanish on physical states and this gives rise to a set of local and global Gauss-law constraints that the quantum fields and their conjugate momenta need to obey. Of these constraints, the local ones can be solved (and will be solved in the next section) and consequently a large subset of the gauge fields' conjugate momenta is written in terms of the fermion color charges. The fact that these momenta become non-dynamical implies that their conjugate gauge fields are not physical and can be removed from the Hamiltonian. These are the fields that, in the action formalism, can be `gauged away'.

Indeed, from the path integral formalism we know that in a compact system of one spatial dimension almost all gauge fields can be gauged away. The only gluonic modes that remain are those corresponding to a constant spatial gauge field. Furthermore, one can gauge away all but the $N-1$ independent eigenvalues of this zero mode. Anticipating a similar scenario in the Hamiltonian approach, one expects that the following fermionic Hamiltonian will be the remnant of the Hamiltonian  \Eq{eq:H_F} after the axial gauge fixing :
\begin{equation}
 H'_F = -i\sum_{x}\psi^{\dag a}_x \sigma_3 \, e^{i\varphi_a } \, \psi^{a}_{x+1}+h.c. + m\sum_x \psi^{\dag a}_x \sigma_1 \, \psi^{a}_x.\label{H'0}
\end{equation}
Here we denote by $e^{i\varphi_a}$ the $a^{\rm th}$ eigenvalues of the constant spatial gauge field operator. 
Clearly, to get \Eq{H'0} from \Eq{eq:H_F} one needs to remove all but the eigenvalues of the spatial Polyakov loops operator from the system. Using a sequence of a change of variables, this is easy to do in the path integral approach. But in the Hamiltonian there seems to be a conceptual difficulty with such a process : the degrees of freedom we wish to gauge away are represented by quantum operators and it is not clear how to `gauge them away'.  

To proceed we choose to generalize, to the lattice, the formalism constructed in the seminal Ref.~\cite{LNT}. The general idea is to find a Hilbert space realization of a unitary operator ${\cal F}$ that will rotate quantum states into a basis where the Hamiltonian looks like \Eq{H'0}. To do so we follow the path integral picture as a guide : first we define the spatial Polyakov loop operator $P$ to be 
\begin{equation}
P_{ab}= \sum_{c,d,\cdots, z=1}^N U_{1,ac} U_{2,cd} \cdots U_{L_s,zb}\,,
\end{equation}
and second we define the `eigenvalue operators' $\varphi_a$ through
\begin{equation}
P_{ab}= \sum_{c} S^\dag_{ac}(P) \, e^{i\,L_s\,\varphi_c(P)} \, S_{cb}(P).\label{P}
\end{equation}
Here the matrix of operators $\left( S(P)\right)_{ab}$ is a functional of the operators $P_{ab}$ and is defined implicitly through \Eq{P}. 
Next we ask to find a form of ${\cal F}$ that will induce the following transformations
\begin{eqnarray}
\psi^a_x &\to& \psi^{'a}_x = {\cal F} \, \psi^a_x \, {\cal F}^{\dag} = \sum_b V_{x,ab} \,\psi^b_x. \label{Psitransform}\\
\psi^{a\dag}_x &\to& \psi^{'\dag a}_x = {\cal F} \, \psi^{\dag a}_x \, {\cal F}^{\dag} = \sum_b \psi^{\dag b}\,V^\dag_{x,ba},\\
U_{x,ab} &\to& U'_{x,ab}={\cal F} \, U_{x,ab} \, {\cal F}^{\dag} = U_{x,ab}, \label{Utransform}\\
U^\dag_{x,ab} &\to& U^{'\dag}_{x,ab}={\cal F} \, U^\dag_{x,ab} \, {\cal F}^{\dag} = U^\dag_{x,ab}.
\end{eqnarray}
Crucially, here the operator $\left(V_x\right)_{ab}$ is chosen to be the following functional of the gauge fields operators $\left(U_{x}\right)_{ab}$,
\begin{eqnarray}
V_{x,ab}&=& \sum_{cde\cdots yz=1}^N U^\dag_{x-1,ac}\, U^\dag_{x-2,cd}\, \dots \, U^\dag_{1,yz} \, S^\dag_{zb}(P) \, e^{i\varphi_b(P) \, x}.\label{Vdef}
\end{eqnarray}
This guarantees that $V$ obeys
\begin{equation}
\sum_{bc}\left(V^\dag_x\right)_{ab} \,\left(U_{x}\right)_{bc}\,\left(V_{x+1}\right)_{cd} = e^{i\varphi_a }\delta_{ad} \label{Vdef1}
\end{equation}
as an operator identity. The form of ${\cal F}$ that induces Eqs.~(\ref{Psitransform})--(\ref{Utransform}) is then written as 
\begin{equation}
{\cal F} = \exp \left( -i\sum_{xi} \rho^i_{x,F}\,  \Theta^i_x(\left\{U\right\})\right),
\end{equation}
where $\Theta$ is again an operator in Hilbert space that depends on the gauge field operators $U_{x,ab}$ and that is defined through its following relation to the operator $V$ :
\begin{eqnarray}
V_{x,ab}&=&\left[\exp \left( i\sum_i \lambda^i \Theta^i_x(\left\{U\right\})\right)\right]_{ab}. \label{Vdef2}
\end{eqnarray}
Note that ${\cal F}$ is {\em not} a gauge transformation (i.e. its not of the form \Eq{gauge_trnsf}). 

The end result of Eqs.~(\ref{Psitransform})--(\ref{Vdef2}) is that when we rotate the Hilbert space by ${\cal F}$ or, equivalently, conjugate the fermionic part of the Hamiltonian by ${\cal F}$ we indeed get \Eq{H'0} :
\begin{eqnarray}
H_F &\to & H'_F = {\cal F} \, H_F \, {\cal F}^{\dag} = -i\sum_{x}\psi^{\dag a}_x \sigma_3 \, e^{i\varphi_a } \, \psi^{a}_{x+1}+h.c. + m\sum_x \psi^{\dag a}_x \sigma_1 \, \psi^{a}_x.\label{H'}
\end{eqnarray}
This change of quantum basis is how the process of axial gauge fixing works in the Hamiltonian formalism.

\bigskip

We now proceed to find the transformed version of $H_E$, of the commutation relations, and of the Gauss-law constraint. We begin by defining the transformed electric fields $E'$ via
\begin{eqnarray}
E^{'i}_x&\equiv & {\cal F} \, E^{i}_x \, {\cal F}^\dag,\label{E'}
\end{eqnarray}
which allows us to write the transformed electric field Hamiltonian as
\begin{eqnarray}
H_E &\to & H'_E={\cal F} \, H_E \, {\cal F}^\dag = \frac{g^2}{2}\sum_{x,i} \left(E^{'i}_x\right)^2.
\end{eqnarray}
It is clear that the commutation relation between the $E'$ fields and the $U$ fields are the same as in \Eq{comm_plus1} and \Eq{comm_plus2} (this is so because of \Eq{Utransform}). Next, we turn to transform the Gauss law constraint. For that we conjugate \Eq{Glaw} by ${\cal F}$
\begin{equation}
{\cal F} \,  \left( E^i_x-\left( U^{\text{Adj.}}_{x-1}\right)_{ji} E^j_{x-1} \right) {\cal F}^\dag = {\cal F} \rho^i_{F,x} \, {\cal F}^\dag.
\end{equation}
This equation, in terms of $E^{'i}_x$, reads
\begin{equation}
E^{'i}_x-\left( U^{\text{Adj.}}_{x-1}\right)_{ji} E^{'j}_{x-1} = {\cal F} \rho^i_{F,x} \, {\cal F}^\dag = \sum_{ab}\, \lambda^i_{ab}\, {\cal F} \, \psi^{\dag a}_x \psi^b_x \, {\cal F}^\dag =  \sum_{ab \atop a'b'}\, \lambda^i_{ab}\,\, \psi^{\dag a'}_x \, \psi^{b'}_x \, V^\dag_{x,a'a} V_{x,bb'}. 
\end{equation}
Multiplying this equation by $\lambda^i_{cd}$, summing over $i$, 
and writing it in terms of the adjoint representation of the operator $V$, we get
\begin{equation}
\sum_{i}\, \left( V^{\rm Adj.}_x \right)_{ik} \,  \left( E^{'i}_{x}-\left( U^{ \text{Adj.}}_{x-1}\right)_{ji} E^{'j}_{x-1} \right) = \rho^k_{F,x}.
\end{equation}
Finally, if we define
\begin{equation}
E^{''k}_x=\left(V^{\rm Adj.}_x\right)_{ik} \, E^{'i}_x, \label{E''def}
\end{equation}
and
\begin{equation}
E^{'j}_x=\left(V^{\rm Adj.}_x\right)_{jk}\, E^{''k}_x,
\end{equation}
then we get
\begin{equation}
 E^{''i}_{x}-\left( V^\dag_x \, U^\dag_{x-1}\, V_{x-1}\right)^{\rm Adj.}_{ij} E^{''j}_{x-1}  = \rho^i_{F,x}.
\end{equation}
Using \Eq{Vdef} this gives
\begin{equation}
 E^{''i}_{x}-\left(e^{-i\varphi}\right)^{\rm Adj.}_{ij} E^{''j}_{x-1}  = \rho^i_{F,x}. \label{Glaw''}
\end{equation}
\Eq{Glaw''} is the  starting point to the discussions in Sections~\ref{GLresolve}--\ref{0mode} : first, in Section~\ref{GLresolve}, we solve \Eq{Glaw''} and write the fields $E^{''i}_x$ in terms of the color charge densities $\rho^i_{F,x}$. This cannot be done to  all of the components of $E^{''}$, and  a subset of these remains independent of the fermionic charge densities. The way we treat these remaining components is explained in Section~\ref{0mode}.   

\section{Resolution of the Gauss law constraint}
\label{GLresolve}

In this section we resolve the Gauss law constraint of \Eq{Glaw''} and write the electric field operators $E^{''}$ in terms of the fermion color charge densities $\rho_F$. For that purpose we first specify the basis of the color group generators $\lambda^i$. We choose the $N-1$ traceless generators $\lambda^{i=1,2,\dots,N-1}$ to span the traceless Cartan sub-algebra, and the remaining $N(N-1)$ generators to have no diagonal entries.  In this basis we see that 
\begin{equation}
\left(e^{-i\varphi}\right)^{\rm Adj.}_{ij} \equiv 2\, e^{i(\varphi_a-\varphi_b)}\, \lambda^i_{ab}\, \lambda^j_{ba}  = \left[ \begin{array}{cc}
\delta_{ij} & {\rm if}\,\,i,j \in [1,N-1],\\
0 & \hspace{0.5cm} {\rm if}\,\, i\in  [1,N-1] {\rm \,\, but\, \, not}\, j, \, \, {\rm and \,vice\,versa}  
\end{array}
\right. .
\end{equation}
As a result, if we focus on the Cartan sub-algebra (let us denote its generators by $i=I$), and Fourier transform the operators $E''$ and $\rho$ according to
\begin{eqnarray}
E^{''I}_x &=& \frac1{\sqrt{L_s}} \, \sum_p \,e^{ipn} E^{''I}_p,\label{FTE''}\\
\rho^{I}_x &=& \frac1{\sqrt{L_s}} \, \sum_p \,e^{ipn} \rho^{I}_p,\label{FTrho},
\end{eqnarray}
with $p=\frac{2\pi n}{L_s}\, ; \, n=0,1,\dots,L_s-1$, then we can resolve the nonzero momentum components of $E^{''I}_{p\neq 0}$ :
\begin{eqnarray}
E^{''I}_{p} &=& \frac{\rho^I_{F,p}}{1-e^{-ip}}, \quad {\rm for} \quad p\neq 0. \label{E''p}
\end{eqnarray}
For $p=0$ Gauss law cannot be resolved, and instead becomes a constraint on the fermionic global color charges :
\begin{eqnarray}
\forall I=1,2,\dots,N-1 \qquad :\qquad Q^I&\equiv& \sum_x\, \psi^\dag_x \, \lambda^I \, \psi_x = 0.\label{QI}
\end{eqnarray}
The $N-1$ requirements of \Eq{QI} can be written as
\begin{equation}
\forall a=1,\dots,N\quad : \quad\sum_x \, \psi^{\dag a}_x \psi^a_x = {\rm independent \,\, of \,\,} a,
\end{equation} 
and since the baryon number $B$ is equal to $\frac1{N}\sum_{x,a} \psi^{\dag a}_x\psi^a_{x}-L_s$ (see \Eq{B}) then \Eq{QI} becomes
\begin{equation}
\sum_x \, \psi^{\dag a}_x \psi^a_x = B+L_s, \qquad \forall a=1,2,\dots,N.\label{BLs} 
\end{equation}

For the rest of the generators $i\neq I$ we can indeed resolve Gauss law even for zero momentum  : multiplying \Eq{Glaw''} by  $\lambda^i_{ab}$ with $a\neq b$, and defining $E^{''}_{x,ab}=\sum_{i\neq I} \, \lambda^i_{ab}\, E^{''i}_x$, and $\rho_{F,x,ab}=\sum_{i\neq I} \, \lambda^i_{ab}\, \rho^{i}_{F,x}$, one finds that the Fourier components of $E^{''}_{x,a\neq b}$ obey 
\begin{equation}
E^{''}_{p,ab} = \frac{\rho_{F,p,ab}}{1-e^{-i(p+\varphi_a - \varphi_b)}}.\label{Glaw''1}
\end{equation}
(here the Fourier transformations of $E^{''}_{x,a\neq b}$ and  $\rho_{x,a\neq b}$ are defined in a way similar to Eqs.~(\ref{FTE''}--\ref{FTrho}))
Note that we work in a Schr\"oedinger picture where $\varphi_a$ is a $c$-number. Also, in \Eq{Glaw''1} we assume the absence of states with $\varphi_a-\varphi_b+p=0$. Since $p$ is quantized in units of $2\pi/L_s$, this assumption will indeed become true -- see the discussion below on the importance of the Jacobian of the curvilinear coordinates $\varphi$.

We can now use the relations between $E^{''}$ and $\rho_F$, and write the electric Hamiltonian in terms of the fermions and the gluonic zero modes. For that we use the hermiticity of $E'_x$ and get 
\begin{eqnarray}
H'_E &=& \frac{g^2}2\sum_{ix} \left( E^{'i}_x \right)^2=\frac{g^2}2\sum_{ix} E^{'i\dag}_x E^{'i}_x=\frac{g^2}2\sum_{ix} E^{''i\dag}_x E^{''i}_x = \frac{g^2}2\sum_{ip} E^{''i\dag}_p\, E^{''i}_p\nonumber \\
&=&\frac{g^2}2\left( \frac2{L_s}\sum_{ab\atop x,y,p} ^\prime \, \frac{\rho_{x,ab}\, \rho_{y,ba}\, e^{i(y-x)\,p}}{4\sin^2 \left(\frac{p+\varphi_a-\varphi_b}{2}\right)} + \sum_{I=1}^{N-1} E^{''I\dag}_{p=0} E^{''I}_{p=0}\right). 
\end{eqnarray}
Here, by the primed sum, we mean that one must sum over all $a$, $b$ and $p$ for which the denominator is {\em nonzero}. This restriction is a direct consequence of one's inability to resolve all of the Gauss law constraints and will become the source of the principle value prescription often used in the `t Hooft model. 

\section{Realization of the zero modes in the Schr\"oedinger picture}
\label{0mode}

The resolution of Gauss law left us in the glue sector with a restricted set of dynamical degrees of freedom -- the $N-1$ eigenvalues of the spatial Polyakov loops, $\varphi_a$, and their $N-1$ conjugate momenta $E^{I''}_{p=0}$. In this section we focus on these operators to which we collectively refer  as the zero modes. Specifically, we calculate their commutation relations and also find a simple way to realize $E^{I''}_{p=0}$ in the Schr\"oedinger picture, where $\varphi_a$ are c-numbers. For that purpose we first calculate the commutation relations of $E^{'i}_x$ and the operator
\begin{equation}
\left(P_y\right)_{ab} = \left( U_1 \, U_2 \cdots U_{y} \right)_{ab}.
\end{equation}
The result is easy to obtain and can be written in terms of the operators $S$ and $\varphi$ (see \Eq{P}), and  the operator $V$ (\Eq{Vdef}), 
\begin{equation}
\left[E^{'i}_x,\left(P_y\right)_{ab}\right] = \theta(y-x) \left( S^\dag \, e^{i\varphi \, x} V^\dag_x \lambda^i \, V_x e^{-i\varphi \, x } S P_y  \right)_{ab}.
\end{equation}
Here $\theta(x)= 1$ if $x\ge 0$ and zero otherwise.
Next we use the definition of $E^{''I}$ in \Eq{E''def} and after some algebra find that (here note that we focus only on the Cartan sub-algebra)
\begin{equation}
\left[E^{''I}_x,\left(P_y\right)_{ab}\right] = \theta(y-x) \left( S^\dag \, \lambda^I \, S P_y  \right)_{ab}.\label{comm1}
\end{equation}
Since the dependence on the site index $x$ in the r.h.s.~of \Eq{comm1} is trivial, we can easily write the commutation relation of the zero mode $E^{''I}_{p=0}=\frac1{\sqrt{L_s}} \, \sum_x E^{''I}_x$, and we find
\begin{equation}
\left[E^{''I}_{p=0},\left(P_y\right)_{ab}\right] = \frac{y}{\sqrt{L_s}} \, \left( S^\dag \, \lambda^I \, S P_y  \right)_{ab}.\label{comm2}
\end{equation}
In particular, for $y=L_s$, we use \Eq{P} and get 
\begin{equation}
\left[E^{''I}_{p=0},P_{ab}\right] = \sqrt{L_s} \, \left( S^\dag \, \lambda^I \, e^{i\varphi L_s} S  \right)_{ab}.\label{comm3}
\end{equation}
If we now define ${\cal E}_a\equiv \sum_I \, \lambda^I_{aa}E^{''I}_{p=0}$, we get 
\begin{eqnarray}
\left[{\cal E}_c,\left(S^\dag e^{i\varphi L_s}S\right)_{ab}\right]&=&\frac12 \, \sqrt{L_s} \left[\, S^{\dag}_{ac} e^{i\varphi_c L_s} S_{cb} -\frac1{N} \, P_{ab}\right].\label{comm4_1}
\end{eqnarray}
In Appendix~\ref{gauge_fixing} we show that 
\begin{eqnarray}
{\cal E}_a&=&{\cal E}^\dag_a,\label{herm}\\
\left[{\cal E}_a,{\cal E}_b\right]&=&0\label{comm5}.
\end{eqnarray}
and we note in passing also that $\sum_a {\cal E}_a=0$.

The form of Eqs.~(\ref{comm4_1})--(\ref{comm5}) leads us to write the following realization of the ${\cal E}$ operators in the Schr\"oedinger picture : 
\begin{equation}
2\sqrt{L_s}\, {\cal E}_a = - i \left( \frac{\delta}{\delta \varphi_a}-\frac1{N} \sum_c \frac{\delta}{\delta \varphi_c}\right) -\frac{i}2 \left[\frac{\delta \, \log \Delta^2 (\varphi)}{\delta \varphi_a} -\frac1{N}\sum_c\frac{\delta\log \, \Delta^2}{\delta \varphi_c}\right],\label{E_Schroedinger0}
\end{equation}
where here $\Delta^2(\varphi)$ is the well known Vandermond determinant
\begin{equation}
\Delta^2(\{\varphi\})=\prod_{a<b}\, \sin^2 \left(L_s\frac{\varphi_a-\varphi_b}{2}\right). \label{VDM}
\end{equation}
It is easy to check that the first term in \Eq{E_Schroedinger0} indeed satisfies \Eq{comm4_1} : the $\varphi_{a=1,\dots,N}$ behave like the coordinates of $N$ particles, while $\frac{\delta}{\delta \varphi_a} - \frac1{N}\sum_b \frac{\delta}{\delta \varphi_b}$ behave like their momenta, relative to the motion of the center of mass coordinate $\varphi_{c.m.}\equiv\sum_a \varphi_a$. This separation into relative coordinates and a center of mass is anticipated in the $SU(N)$ case where $\varphi_{c.m.}$ is not a true degree of freedom.

It is also not hard to understand the origin of the second term in \Eq{E_Schroedinger0} : while it trivially obeys \Eq{comm4_1}, it is necessary for \Eq{herm}. To see this recall that in the Schr\"oedinger picture the measure of the $\varphi_a$ coordinates is not flat -- it is given by the Haar measure $dP$ over the spatial Polyakov loop $P$ (which is the only remnant of the gauge field degrees of freedom). Since the Hamiltonian only depends on the eigenvalues $\varphi_a L_s$ of $P$, then we can replace 
\begin{equation}
\int_{SU(N)} dP\longrightarrow 
\int \, \left(\prod_a d\varphi_a\right)\, \,\Delta^2(\left\{\varphi\right\})\, \delta \left(\varphi_{c.m.} \right).
\end{equation}
Because of the $\Delta^2$ factor in the measure, the simple derivative operator $-i\frac{\delta}{\delta\varphi}$ will not be hermitian (its action to the right will differ from its action to the left by a derivative of $\Delta^2$). It is easy to check the particular choice of in \Eq{E_Schroedinger0} fixes this and makes \Eq{E_Schroedinger0} hermitian.\footnote{The integral over $\varphi$ can be either restricted to obey $\varphi_a \ge \varphi_{b}$ for $a>b$, or can be simply unrestricted. The difference between these two choices will be reflected by whether the operators $S(P)$ include permutations of the eigenvalues or not.}

Our final step will be to simplify \Eq{E_Schroedinger0} even further. For that we define a new wave function $\Psi_{\rm new}$ for the curvilinear coordinates $\varphi$ that will make their measure flat. Specifically, we write
\begin{equation}
\Psi(\varphi)\equiv \frac{\Psi_{\rm new}(\varphi)}{\prod_{a<b} \, \sin\left( L_s \frac{\varphi_a-\varphi_b}2 \right)}.\label{change}
\end{equation}
In terms of the new wave function the Vandermond disappears from the measure and the kinetic term becomes the simple quadratic form
(here we rescaled $\varphi \to \varphi/L_s$):
\begin{eqnarray}
\frac{g^2}{2}\sum_I \left(E^{''I}_{p=0}\right)^2 &=& -\frac{g^2L_s}{4} \sum_{d=1}^N\left( \frac{\delta}{\delta\varphi_d}-\frac1{N}\sum_c \frac{\delta}{\delta\varphi_c} \right)^2.
\end{eqnarray}
This will be our final simplified form for the zero mode contribution to $H^{'}_E$.

\section{Recap : the Hamiltonian and the restrictions on the Hilbert space.
}
\label{Hrecap}

To conclude let us write the lattice Hamiltonian $H$ in terms of the operators that cannot be gauged away. It describes the interactions between the lattice fermions $\psi$ and the eigenvalues of the spatial Polyakov loop $\varphi$, and is given by
\begin{eqnarray}
H&=&H_G + H_K + H_C,\label{Ham}\\
H_G&=&-\frac{g^2L_s}{4} \sum_{d=1}^N\left( \frac{\delta}{\delta\varphi_d}-\frac1{N}\sum_{c=1}^N \frac{\delta}{\delta\varphi_c} \right)^2,\label{HG}\\
H_K & =& -\frac{i}2\sum_{x}\psi^{\dag a}_x \, e^{i\varphi_a/L_s} \, \psi^{a}_{x+1}+h.c. + m\sum_x \, (-1)^x\, \psi^{\dag a}_x  \psi^{a}_x,\label{H_F'}\\
H_C&=&\frac{g^2}{L_s}\sum'_{abp\atop}\,\frac{\rho^{ab}_{F,x} \,\rho^{ba}_{F,y}\,e^{ip(y-x)}}{4\sin^2\, \left(\frac{(\varphi_a-\varphi_b)/L_s+p}{2}\right)},\label{HC}\\
\rho^{ab}_{F,x} &\equiv & \frac12 \left(\psi^{\dag b}_x\psi^a_x - \frac{\delta^{ab}}N\sum_c \psi^{\dag c}_x\psi^c_x  \right).
\end{eqnarray}
Here the primed sum in \Eq{HC} means we that we should not sum over terms whose denominator is zero (for the derivation of this restriction see section \ref{GLresolve}). From here on we discard the lower component of the fermions, and so discuss the staggered fermions formulation of lattice QCD \cite{Susskind} (this also means that we need to replace the r.h.s. of \Eq{BLs} by $B+L_s/2$, and choose the number of lattice sites, $L_s$, to be even). The reason we choose to work with staggered fermions is two-fold. Firstly, these one-component Grassmann variables are simpler to treat in our formalism and the computational effort involved in the numerical minimization of their classical Hamiltonian (for details see \cite{nonzeroBpaper}) is a factor of two smaller than for the naive prescription of fermions considered above. Secondly, unlike the situation in four dimensional euclidean calculations, staggered fermions in the one dimensional Hamiltonian approach are `un-doubled' and correspond to a single Dirac fermion in the continuum limit. 

\bigskip
Finally, the Hilbert space of the system obeys the following : 
\begin{itemize}
\item The operators $\sum_{x}\, \psi^{\dag a}_x \psi^a_x$ are, for each value of the color index $a=1,\dots,N$, constants and equal to $B+L_s/2$.
\end{itemize}

\section{Diagonalization of the Hamiltonian : the coherent states approach}
\label{diagonalize}

In this section we use the coherent state approach to study the ground state properties of the Hamiltonian in its large-$N$ limit. This approach is summarized in Refs.~\cite{YaffeCoherent} and we refer to these papers for detailed discussions, while here we only describe its strategy. 

\bigskip
The main paradigm that underlies the coherent state approach is that QCD, in its large-$N$ limit, becomes a classical theory.\footnote{Here we implicitly assume that we take the large-$N$ limit when all other parameters are fixed. These includes the volume $L$, the temperature $T$, the mass $m_q$, the `t Hooft coupling $g^2N$, and the baryon number $B$.} This means that instead of diagonalizing its quantum Hamiltonian, one can instead minimize a corresponding classical function ${\cal H}$, referred to as the  `classical Hamiltonian'. The minimization is done with respect to a set of coordinates, ${\cal C}$, that corresponds to expectation values of gauge invariant operators. The coherent state approach provides the mathematical prescription for calculating the function ${\cal H}({\cal C})$ and it is shown to be given by
\begin{equation}
{\cal H}({\cal C}) = \<{\cal C}|H|{\cal C}\>.
\end{equation}
Here $|{\cal C}\>$ is the so called `coherent state' obtained by applying a unitary color singlet operator, that we denote by ${\cal U(C)}$, on an arbitrary color singlet element $|0\>$ of Hilbert space, the so called `reference' state :
\begin{equation}
|{\cal C}\>={\cal U(C)} |0\>. \label{CS}
\end{equation}
The operator ${\cal U}$ is a functional of gauge invariant operators like spatial Wilson loops of contour $\Gamma$
\begin{equation}
\tr \, W_\Gamma,\label{W}
\end{equation}
as well as spatial Wilson loops decorated by a single insertion of the electric field matrix operator $\displaystyle{\left(E_x\right)_{ab}\equiv \sum_{i=1}^{N^2-1} E^i\, \lambda_{ab}}$, at point $x$ along the loop, i.e.
\begin{equation}
\tr \, E_{x\in \Gamma} \, W_{\Gamma}, \quad {\rm and} \quad  \tr W_{\Gamma} \, E_{x\in \Gamma}.\label{Wdecor}
\end{equation}
(the two operators that appear in \Eq{Wdecor} are different since the electric fields do not commute with the gauge fields). Also, in the presence of fermions, ${\cal U}$ depends on string-like operators of the form  
\begin{equation}
\psi^\dag_x \, U_{x\to y} \psi_y,\label{string}
\end{equation}
where here $U_{x\to y}$ is a string of gauge field operators connecting site $x$ with site $y$. The generic way ${\cal U}({\cal C})$ depends on these operators is 
\begin{eqnarray}
{\cal U}({\cal C}) &\equiv& {\cal U}_F\times {\cal U}_G,\label{Udef1}\\
{\cal U}_F &\equiv& \exp\, \left[ \sum_{xy} {\cal C}^{xy}_F\, \psi^\dag_x \, U_{x\to y} \psi_y \right],\\
{\cal U}_G &\equiv& \exp\, \left[ \sum_\Gamma \, {\cal C}^\Gamma_G\,\tr \, W_\Gamma +  \sum_{\Gamma}\sum_{x\in \Gamma} \, \left({\cal C}^{x,\Gamma}_{G,1}\,\tr \, E_{x}\, W_\Gamma  +  {\cal C}^{x,\Gamma}_{G,2}\,\tr \, W_\Gamma E_{x}\right)\right],\label{Udef3}
\end{eqnarray} 
where  ${\cal C}^{xy}_F,{\cal C}^{\Gamma}_G$, and ${\cal C}^{x\Gamma}_{G,1,2}$ are $c$-numbers.

Since all the three type of operators defined in Eqs.~(\ref{W}--\ref{string}) form a closed algebra, the structure of Eqs.~(\ref{Udef1}--\ref{Udef3}) guarantees that ${\cal U}({\cal C})$ is an element of a Lie group referred to as the coherence group. Indeed ${\cal C}^{xy}_F,{\cal C}^{\Gamma}_G$, and ${\cal C}^{x\Gamma}_{G,1,2}$ parameterize this Lie group and furnish coordinates on its manifold. More precisely, the fermionic part of the coherence group is parameterized by the fermionic coordinates ${\cal C}_F$, and the gluonic part by ${\cal C}_{G}$ and ${\cal C}_{G,1,2}$.\footnote{To ensure that ${\cal U}(\cal C)$ is unitary, the coordinates ${\cal C}$ need to obey certain conditions and in the next section we make these explicit.} 

\bigskip

The values ${\cal C}_{\rm min}$ that minimize ${\cal H(C)}$ then determine the values of all gauge invariant observables in the ground state of $H$, so for example, the ground state energy $E_{\rm g.s.}$ is given by
\begin{equation}
\lim_{N\to \infty} E_{\rm g.s.} =  {\cal H(C=C_{\rm min})},
\end{equation}
etc.

\bigskip
In the large-$N$ limit of QCD with the number of flavors, $N_f$, kept fixed, the minimization of ${\cal H}$ process proceeds in two steps : one begins by minimizing the leading $O(N^2)$ contribution of ${\cal H}({\cal C})$. At leading order the latter is equal to the classical function ${\cal H}_G$ given by the classical Hamiltonian of the pure gauge theory 
\begin{equation}
{\cal H}_G = \<0_g|\, {\cal U}^\dag_G\, H_G \, {\cal U}_G\, |0_g\>,
\end{equation}
where the operator $H_G$ is given in \Eq{HG} and $|0_g\>$ is a reference state in the pure gauge system. This step sets the value of the expectation values of the gluonic color singlet operators such as the spatial Polyakov loops. For brevity, let us refer to these expectation values by the generic symbol ${\cal P}$, and by ${\cal P}_{\rm min}$ to their value at the minimum of ${\cal H}_G$.

Next, one  minimizes the $O(N)$ contribution to ${\cal H}$. To leading order it is given by calculating the expectation value of $H_K+H_C$ from Eqs.~(\ref{H_F'}) and (\ref{HC}) in the subset of coherent states whose gluonic coordinates have already been determined by the gauge dynamics. This contribution is thus given by 
\begin{equation}
{\cal H}_F({\cal C}_F)\equiv \<0_F|\, {\cal U}^\dag_F\, \left(H_K+H_C\right)\, {\cal U}_F\, |0_F\>_{|{\cal P}={\cal P}_{\rm min}},
\end{equation}
with $|0_F\>$ denoting a fermionic reference state.
 This two step process reflects the dominance of the gauge field dynamics over the fermion dynamics. Indeed the back-reaction of the fermions on the gauge fields comes from terms which we do not consider in this work and that are subleading in $1/N$. This is true as long as there is no enhancement of these $1/N$ terms by massless modes, which we assume to be the case.\footnote{We thank V.~P.~Nair for pointing this to us and to L.~G.~Yaffe for a discussion related to this.} 

For our purposes it will be easier to simply calculate ${\cal H}_F({\cal C}_F)$ and substitute the set of expectation values ${\cal P}$ by its value in the {\em exact} ground state of $H_G$. The latter is known analytically and we repeat its derivation in Section~\ref{SectorG}. The solution of the fermion sector, however, is done as described above and we calculate the classical function ${\cal H}_{F}({\cal C}_F)$ in Section~\ref{SectorF}. 

\bigskip

We now turn to make the following remark on the way ${\cal H}$ is calculated and minimized. As mentioned above, the coordinates ${\cal C}$ corresponds to expectation values of different color singlet operators. More accurately, one can focus ones attentions to expectation values of only `non-factorizable' operators such as the string operators of \Eq{string}.
Naively, however, one might expect that ${\cal H}$ should also be minimized with respect to coordinates that correspond to expectation values of `factorizable' operators such as
\begin{equation}
\frac1{N} \left( \psi^\dag_x \psi_x \right)\,\times\, \left(  \psi^\dag_y \psi_y \right), \frac1{N}\left(\psi^\dag_x \, U_{x\to y} \psi_y \right)\,\times \, \left(\psi^\dag_z \,\psi_z\right), \dots.
\end{equation}
This, however, is incorrect : at large-$N$ the expectation values of such operators factorize, and for example,
\begin{equation}
\<\psi^\dag_x \psi_x \,\times \, \psi^\dag_y \psi_y \> \stackrel{N=\infty}{=} \< \psi^\dag_x \psi_x \>\,\times\, \<  \psi^\dag_y \psi_y \>.
\end{equation}
As a result, the expectation values of factorizable operators become determined by the expectation values of the non-factorizable operators and should not be thought of as independent coordinates that parameterize the coherent state manifold or the point within that manifold that represents the ground state.

The unique role of non-factorizable operators operators is also reflected by the structure of the unitary operator of Eqs.~(\ref{Udef1}--\ref{Udef3}) which depends only on such operators. Indeed in Ref.~\cite{YaffeCoherent} it is shown that to generate the whole coherence group, it is sufficient to include in its algebra only the non-factorizable operators that appear in Eqs.~(\ref{Udef1}--\ref{Udef3}). Adding other operators greatly complicates the algebra of the coherence group and is unnecessary. 
 An important result of the discussion above, which we wish to emphasize, is that the coherent state $|{\cal C}\>$ defined in \Eq{CS} is sufficiently general to look for all possible large-$N$ ground states. Put differently, the form in \Eq{CS} does {\em not} correspond to assuming an ansatz for the ground state of the gauge theory.

\bigskip

In the next two subsections we show how to implement the program outlined above in practice. We begin in Section~\ref{SectorG} with the treatment of the pure gauge case that allows us to calculate the expectation values of the color singlet gluonic operators, namely the traces of different powers of spatial Polyakov loops. The pure gauge Hamiltonian in our two-dimensional case is sufficiently simple that we can do so exactly and for any value of $N$. The next step is to apply the coherent state approach to the fermionic part of the Hamiltonian and we do so in Section~\ref{SectorF}.

\subsection{Solution of the gauge sector}
\label{SectorG}
The solution of the pure gauge sector is well known \cite{Douglas}. The starting point is to notice that \Eq{change} means that the gauge wave functions need to be anti-symmetric to an exchange of two angles, and for odd values of $N$, periodic in $2\pi$:
\begin{eqnarray}
\Psi(\varphi_1,\varphi_2,\dots,\varphi_k,\dots,\varphi_l,\dots,\varphi_N) &=& -\Psi(\varphi_1,\varphi_2,\dots,\varphi_l,\dots,\varphi_k,\dots,\varphi_N).\\
\Psi(\varphi_1,\varphi_2,\dots,\varphi_k+2\pi,\dots,\varphi_N) &=& +\Psi(\varphi_1,\varphi_2,\dots,\varphi_k,\dots,\varphi_N).
\end{eqnarray}
These properties, together with the form of the Hamiltonian, tell us to think of the $\varphi_a$ as the positions of $N$ non-relativistic fermions with mass $2/g^2L_s$, moving on a circle with periodic boundary conditions and a fixed center of mass. The single particle wave functions of such a system are the plane waves $e^{i\varphi n}$ with $n=0,\pm 1,\pm 2,\dots$, and the ground state wave function of this $N$-fermion system is, up to a phase, the $N\times N$ slater determinant obtained by occupying momentum states distributed symmetrically around zero and limited by `Fermi momenta' $n_F=(N-1)/2$:
\begin{equation}
\Psi(\{\varphi\}) = \det_{-{n_F} \le a,b \le +n_F} e^{i \, b\,\varphi_a }. \label{gaugeGS}
\end{equation}
This wave function \Eq{gaugeGS} is an eigenstate of $H_G$ with eigen-energy that has the large-$N$ limit of $\frac{L_sN^2}{48}\times g^2N$. 

It is easy to show that this determinant differs from the Vandermond determinant \Eq{VDM} by a phase that depends only on the `center of mass' of the fermions, $\sum_c\varphi_c$ , and in $SU(N)$ we set this phase to zero. The result is
\begin{equation}
|\Psi(\{\varphi\})|^2 = \Delta^2(\{\varphi\}),
 \label{gaugeGS1}
\end{equation}
and so when we calculate expectation values of gauge invariant gluonic operators $O(\left\{\varphi\right\})$ we need to perform the following integral (here we set the normalization such that $\<1\>=1$ and denote $\sum_c \varphi_c$ by $\varphi_{c.m.}$)
\begin{eqnarray}
\<\hat O\>_{G} &=& \int d\varphi \, \Delta^2(\{\varphi\})\,\delta\left(\varphi_{c.m.} \right)\, \hat O(\left\{\varphi\right\}),
 \label{vevG}\\
\int d\varphi&\equiv& \frac1{N!}\int \, \prod_{a=1}^N \frac{d\varphi_a}{2\pi}.
\end{eqnarray}

\subsection{Solution of the fermion sector : large-$N$ coherent states}
\label{SectorF}

We diagonalize the fermion sector in a variational manner. As discussed above, the most general form of the coherent state $|{\cal C}_F\>$ is given by 
\begin{equation}
|{\cal C}_F\> \equiv {\cal U}({\cal C}_F) |0\> = \exp \left( -i \sum_{x \in Z_{L_s}}\sum_{ y \in Z} \, {\cal C}^{xy}_F \, \psi^{\dag \, a}_x \, e^{i\varphi_a(y-x)/L_s} \, \psi^a_y\right) \, |0\>, \label{theta}
\end{equation}
and parameterized by the infinite dimensional hermitian matrix ${\cal C}_F$.  
The state $|0\>$ is a fermionic reference state which we choose to be a state annihilated by $\psi^\dag_x$ for a subset of lattice sites that are full of fermions and that we denote by $S$. More precisely $|0\>$ is defined to obey
\begin{equation}
\psi^\dag_{x,a} |0\> =0 \quad ;\quad \forall x\in S.
\end{equation}
 On the complementary set of sites, $x\in \bar S$, the reference state is annihilated by $\psi_{x}$. This means that we choose $B+L_s/2$ of the lattice sites to be full of baryons and the rest empty of baryons. This choice is convenient for our calculation, but the results are insensitive to it : the only thing that matters is that the overall baryon number that $|0\>$ contains is $B+L_s/2$ (such that the renormalized baryon number is $B$). 

We emphasize here that in the exponent of \Eq{theta}, the index $x$ runs over $1,2,\dots, L_s$, while the sum over $y$ is unrestricted : $y=0,\pm 1,\pm 2,\dots, \pm \infty$. The identification 
\begin{equation}
\psi^a_x=\psi^a_{x+k\,L_s} \quad ;\quad k\in Z,
\end{equation}
means that ${\cal C}_F$ needs to obey
\begin{equation}
{\cal C}^{xy}_F={\cal C}^{x+k\,L_s,y+k\, L_s}_F\quad ;\quad k\in Z, \label{thetaBC}
\end{equation}
in order to have ${\cal U}({\cal C}_F)\,{\cal U}^\dag({\cal C}_F)={\bm 1}$.

\bigskip

To find what is ${\cal C}_F$ we minimize the classical Hamiltonian ${\cal H}$ defined by
\begin{equation}
  {\cal H}_F\equiv \int d\varphi\,\, \<{\cal C}_F|H_K+H_C|{\cal C}_F\> \, \Delta^2(\varphi)\times  \delta\left(\varphi_{c.m.}\right),\label{Hclassical}
\end{equation}
in the space of all possible choices of ${\cal C}_F$.
The calculation of $\<{\cal C}_F|H_K+H_K|{\cal C}_F\>$ is somewhat lengthy and we postpone it to Appendix~\ref{appHF}.

The resulting expression can be written in terms of the following `density matrices' $\rho^q_{xy}$ and $\bar \rho^q_{xy}$ with $x,y\in [1,L_s]$ and $q\in Z$
\begin{eqnarray}
\rho^q_{xy}&\equiv&\sum_{p\in Z\atop z\in S}\, \left(e^{i{\cal C}_F}\right)^{z+pL_s,x}\, \left(e^{-i{\cal C}_F}\right)^{y,z+(q+p)L_s},\label{rho}\\
\bar \rho^q_{xy}&\equiv& \delta_{q,0} \delta_{xy} - \rho^q_{xy},\label{rhobar}
\end{eqnarray}
In terms of $\rho^q_{x,y}$ and $\bar \rho^q_{x,y}$ the classical Hamiltonian is 
\begin{eqnarray}
\<{\cal C}_F|H_K+H_C|{\cal C}_F\> &=& \sum_{x\in Z_{L_s}\atop q \in Z} \left[ \left(-\frac{i}2 \rho^q_{x,x+1}  + c.c. \right)+ m (-1)^x \rho^q_{xx} \right]\times \sum_a e^{i\varphi_a}\nonumber \\
&+&\frac{g^2}{4L_s} \sum_{xy\in Z_{L_s}}\sum_{p\atop a\neq b} \sum_{qq'\in Z} \frac{\rho^{q'}_{xy}\, \bar \rho^{-q}_{yx} \, e^{-i(x-y)p }\times  e^{-i\varphi_a(x-y +q L_s)/L_s + i\varphi_b( x-y + q'L_s)/L_s}}{4\sin^2((\varphi_a-\varphi_b)/L_s + p)/2}  + O(1/N). \nonumber \\\label{HKC}
\end{eqnarray}
Note that in contrast to the form \Eq{HC}, here the second term does not contain any contributions from the $a=b$ terms. They are shown to be subleading in Appendix~\ref{appHF}.

To obtain a compact form for the classical Hamiltonian ${\cal H}_C$ we define $P_k$ to be the Polyakov loop operator that winds $k$ times around the torus as 
\begin{equation}
P_k \equiv \frac1{N} \sum_a e^{i\varphi_a k},
\end{equation}
and use the identity 
\begin{equation}
\frac1{4\sin^2\left((k+i\epsilon)/2\right)} = - \frac12 \sum_{Q\in Z} \, |Q|\, e^{-\epsilon |Q| + iQk},\label{FT}
\end{equation}
with $\epsilon>0$ regularizing the pole of the left hand side, to re-write the second term of \Eq{HKC} as a sum over terms that only contain powers of $e^{i(\varphi_a-\varphi_b)/L_s}$. Indeed, substituting \Eq{FT} into the second term of \Eq{HKC}, we find that the terms contributing to the  sum over the momentum variable $p=2\pi\, l/L_s$ can be isolated and read
\begin{equation}
\sum_{l=0}^{L_s-1} e^{-i2\pi l(x-y - Q)/L_s} = L_s \, \sum_{\bar Q \in Z}\delta_{y-x+Q,\bar QL_s}.
\end{equation}
This allows us to substitute $Q$ in \Eq{FT} by  $x-y+\bar QL_s$. Since the dummy summation variable $\bar Q$ obtains all possible integer values, we drop the `bar' from its notation in the rest of the paper. Thus, the final result of the manipulations in the last paragraph is that \Eq{HKC} gets the following form.
\begin{eqnarray}
{\cal H}_F(\rho)/N &=& \sum_{x\in Z_{L_s}\atop q \in Z} \left[ \left(-\frac{i}2 \rho^q_{x,x+1}  + c.c. \right)+ m(-1)^x \rho^q_{xx} \right] \<P_q\>_G \nonumber \\
&-&\frac{g^2N}{8} \sum_{xy\in Z_{L_s}}\sum_{Qqq'\in Z} \rho^{q'}_{xy}\, \bar \rho^{-q}_{yx} \, |x-y+QL_s| e^{-\epsilon|x-y+QL_s|} \left( \<P_{Q-q} P_{-(Q-q')}\>_G - \frac1N \<P_{q'-q}\>_G\right), \nonumber \\\label{HKC1}
\end{eqnarray}
where here we define the gluonic exception values $\<,\>_G$ in \Eq{vevG}.

\subsubsection{Remarks on the form of the classical Hamiltonian}
\label{remarks}

\Eq{HKC1} above is quite an important ingredient in our paper and so let us now pause here and make the following remark on its form and its implications. What \Eq{HKC1} tells us is that the fermionic properties of the system, which are represented by the density matrices $\rho^q_{xy}$, self interact as well as couple to the Polyakov loops $P_q$. The expectation values of the latter are determined by the gauge dynamics and feel no back reaction from the fermions. Thus, performing the minimization in Section~\ref{SectorG} correctly, and determining the properties of the gluonic vacuum in a consistent way, is crucial to get the correct fermion dynamics. For example, in the Sections~\ref{otherworks} and~\ref{decompact} we emphasize that mistreating the glue sector, which is what one does when one ignores the zero modes, leads, through the way \Eq{HKC1} couples the glue sector to the fermion sector, to erroneous results for various fermionic expectation values.

In particular, one can see that if the expectation values of the different Polyakov loops are incorrectly chosen to be unity,
\begin{equation}
P_q =1, \qquad \forall q,
\end{equation}
 then by using large-$N$ factorization of the double trace term $\<P_{Q-q}\,P_{-(Q-q')}\>\stackrel{N=\infty}{=}\<P_{Q-q}\>\,\<P_{-(Q-q')}\>$, one finds that the second term in \Eq{HKC1} is strongly dependent on $L_s$ (this is shown explicitly in Section~\ref{symm}). This volume dependence is contradicting general arguments on large-$N$ gauge theories such as those of Ref.~\cite{EK}.

\subsubsection{Further manipulations of the classical Hamiltonian and preparing for its minimization}
\label{more}

In this section we make further important simplifications of \Eq{HKC1} that will also allow us to minimize it (see for example Ref.~\cite{nonzeroBpaper} and the next sections).

\bigskip

For a $U(N)$ theory one drops the delta functions in \Eq{vevG}, and the gluonic expectation values $\<P_q\>_G$ and $\<P_q\, P_{q'}\>_G$ are calculated in Appendix~\ref{appHF} (see Eqs.~\ref{Pk}--\ref{PkPk'}). In the $SU(N)$ case the integral in \Eq{vevG} changes only by restricting the sum $\sum \varphi_a$ to be zero, and so we expect the Polyakov loops to be the same in the large-$N$ limit (we show this explicitly for $\<P_q\>_G$ in Appendix~\ref{appHF}). Using Eqs.~(\ref{Pk})--(\ref{PkPk'}) we see that, as expected, the leading contribution to \Eq{HKC1} is from the $q=0$ term in the first line and from the $q=q'=Q$ in the first term of the second line. This gives us\footnote{According to Eqs.~(\ref{Pk})--(\ref{PkPk'}) the terms with $q=q'\neq Q$ are also nonzero, but using the representation in Eqs.~(\ref{constraints_res1})--(\ref{constraints_res2}) we verified that they are subleading in $N$ if we take $M\to\infty$ with or after we take $N\to \infty$.}
\begin{eqnarray}
{\cal H}_F(\rho)/N &=& \sum_{x\in Z_{L_s}} \left[\left(-\frac{i}2 \rho^0_{x,x+1}  + c.c. \right)+ m(-1)^x \rho^0_{xx}  \right]\nonumber \\
&-&\frac{g^2N}{8} \sum_{xy\in Z_{L_s}}\sum_{Q\in Z} \rho^{Q}_{xy}\, \bar \rho^{-Q}_{yx} \, |x-y+QL_s| e^{-\epsilon|x-y+QL_s|}.\nonumber \\\label{HKC11}
\end{eqnarray}

Since the index $Q$ runs over all integers we can write
\begin{equation}
\rho^Q_{xy}=\int_{-\pi}^{\pi} \frac{dp}{2\pi} \rho_{xy}(p) e^{ipQ}, \label{FT2}
\end{equation} 
which, re-using the identity \Eq{FT} gives the form
\begin{eqnarray}
{\cal H}_F(\rho)/N &=& \int \frac{dp}{2\pi}\sum_{x\in Z_{L_s}} \left\{  \left(-\frac{i}2 \rho_{x,x+1}(p)  + c.c. \right)+ m(-1)^x \rho_{xx}(p)\right\}  \nonumber \\
&+& \frac{g^2N}{4}  \int \int \frac{dp}{2\pi} \frac{dp'}{2\pi}  \,\frac1{L_s}\,\sum_{xy\in Z_{L_s}}\sum_{l=1}^{L_s} \frac{\rho_{xy}(p)\, \bar \rho_{yx}(p') \, e^{i2\pi l(x-y)/L_s}}{4\sin^2\left((p-p')/L_s+2\pi l/L_s\right)/2} .\nonumber \\\label{HKC2}
\end{eqnarray}

The pole at $p=p'$ and $l=L_s$ in \Eq{HKC2} might seem alarming and tracking back its source to \Eq{HKC} one finds that it is the double sum over the color indices $a$ and $b$ that appears in the second term there. Specifically, terms in that sum for which $|a-b|$ is small are causing this divergence. For these terms the argument of the sine in the denominator can be small; for example if $p=2\pi$ and the Polyakov loops have zero expectation value in the gluonic ground state, then heuristically $\varphi_a\sim 2\pi a/N$. (This is not a gauge invariant statement, but for the current discussion this subtlety is not important. In the rest of this paper we take great care to avoid such statements when it is important to do so.) Thus, we see that when $|a-b|\stackrel{<}{_\sim} O(1)$, then  the argument of the sine in \Eq{HKC} is very small and this corresponds to the pole in \Eq{HKC2} above.

This pole is the source of the IR divergence in the usual treatments of the `t Hooft model, and that is usually resolved with the {\em ad hoc} principle value prescription. In our case this is not neede. The $a=b$ terms were excluded from the sum in \Eq{HKC}. The way this restriction emerges in \Eq{HKC2} is through certain conditions obeyed by the density matrices $\rho^Q_{xy}$ and $\bar \rho^Q_{xy}$ which we present in Appendix~\ref{appHF} (see Eqs.~(\ref{r11})--(\ref{r21})) and that reflect the unitarity of the operator ${\cal U}({\cal C}_F)$. In the language of $\rho_{xy}(p)$ these conditions read
\begin{equation}
\sum_{y\in Z_{L_s}} \rho_{xy}(p) \bar \rho_{yz}(p) = 0.
\end{equation}
Thus the divergence in \Eq{HKC2} is removed and we get the principle value prescription. To show that the near vicinity of $p=p'$ (corresponding to $|a-b|$ small but nonzero) is not causing any lower divergences we need to assume a form for the $p$ dependence of $\rho_{xy}(p)$. Instead we have confirmed this numerically whilst minimizing ${\cal H}(\rho)$ with respect to $\rho$ \cite{nonzeroBpaper}.

\bigskip

We conclude this section by writing \Eq{HKC2} in a way which is convenient for its minimization. We first solve the constraints Eqs.~(\ref{r11})--(\ref{r21}) on $\rho^Q_{xy}$ by writing 
\begin{eqnarray}
\rho^Q_{xy}&\equiv & \frac1{M} \sum_{a=1}^M \,\sum_{n=1}^{B+L_s/2} \, \phi^n_a(x)\,e^{\frac{2\pi iQa}{N}}\, \phi^{n\star}_a(y), \label{constraints_res1}\\
\bar \rho^Q_{xy}&\equiv & \frac1{M} \sum_{a=1}^M \,\sum_{n=B+L_s/2+1}^{L_s} \, \phi^n_a(x)\,e^{\frac{2\pi iQa}{N}}\, \phi^{n\star}_a(y), \label{constraints_res2}
\end{eqnarray}
where for each $a=1,\dots,M$, the single particle wave functions $\phi^n_a(x) \, ; \, n\in [1,L_s]$ span an $L_s\times L_s$ dimensional space,
\begin{equation}
\sum_{x\in Z_{L_s}} \phi^n_a(x)\, \phi^{m\star}_a(x) = \delta_{mn}.\label{ortho}
\end{equation}
\Eq{constraints_res1} is nothing but a discretized way to write the most general expression for $\rho$ and $\bar \rho$, that also obeys Eqs.~(\ref{r11})--(\ref{r21}) and we show that this is correct in Appendix~\ref{app_rho_resolve}. Note that the full space of solutions for $\rho$ and $\bar\rho$ is accessible only if $M=\infty$.

In terms of $\phi^n_a(x)$ we get
\begin{eqnarray}
{\cal H}_F(\phi)/N &=&  \frac1{M} \sum_a\sum_{x\in Z_{L_s}}\left\{  \left(-\frac{i}2 \rho^a_{x,x+1}  + c.c. \right)+ m (-1)^x \rho^a_{xx}\right\}  \nonumber \\
&-& \frac{g^2N}{4}  \frac1{L_sM^2} \sum'_{abl} \sum_{xy\in Z_{L_s}} \frac{\rho^a_{xy}\, \rho^b_{yx} \, e^{-i(x-y)\left(\frac{2\pi}{L_s}(\frac{a-b}{M}+l)\right)}}{4\sin^2\left(\frac{2\pi}{L_s}(\frac{a-b}{M}+l)/2\right)}\nonumber \\ \nonumber \\
&+& \frac{g^2N(B+L_s/2)}{4}  \frac1{L_sM^2} \sum'_{abl} \frac1{4\sin^2\left(\frac{2\pi}{L_s}(\frac{a-b}{M}+l)/2\right)},\label{HKC4}\label{HF_again} \\
\rho^a_{ab}&\equiv&\sum_{n=1}^{B+L_s/2}\, \phi^n_a(x)\, \phi^{n\star}_a(y).
\end{eqnarray}
Here, by the prime on the sums we mean that the terms with $a=b$ and $l=L_s$ are excluded.\footnote{In our numerical studies we find that, while irrelevant for the minimization of ${\cal H}_F$, the last term in \Eq{HF_again} is crucial to include in order to get the right baryon mass.} 

\bigskip

 We now perform the variation of ${\cal H}$ with respect to the $M$ functions $\phi_a$ 
\begin{equation}
\frac{\delta}{\delta (\phi^n_a(x))^\star} \left( {\cal H} - \sum_{m,b} \epsilon^b_m \sum_x \phi^m_b(x) \, \phi^{m\star}_b(x) \right)=0, \label{variation}
\end{equation}
and we find that they must obey the following $M$ coupled nonlinear differential equations (here we use the Lagrange multiplier $\epsilon^a_{n}$ to enforce \Eq{ortho})
\begin{eqnarray}
\sum_{y\in Z_{L_s}} \, h^a_{xy} \ \phi^n_a(y) &=& \epsilon^a_n \, \phi^n_a(x),\label{diagEQ}
\end{eqnarray}
with
\begin{eqnarray}
h^a_{xy}&=& +\frac{i}2 \left(\delta_{y,x+1} - \delta_{y,x-1}\right) + m\, (-1)^x\, \delta_{xy} -g^2N \, v^a_{xy}, \\
v^a_{xy}&=&\frac1{2M}\sum_b \, K_{ab}(y,x)\, \left(\sum_{m=1}^{B+L_s/2}\, \phi^m_b(x)\, \phi^{m\star}_b(y)\right).
\end{eqnarray}
and
\begin{eqnarray}
K_{ab}(y,x) &=&  \frac1{L_s}\,\sum_{l\in Z_{L_s}}' \, \frac{e^{\frac{2\pi i(x-y)}{L_s}\left(\frac{a-b}{M}+l\right)}}{4\sin^2\left(\frac12\left(\frac{2\pi (a-b)}{ML_s} + \frac{2\pi l}{L_s}\right)\right)}.\label{Kernel}
\end{eqnarray}
Since $K_{ab}$ explicitly depends on $\phi$ then the solution of \Eq{variation} is a self-consistent process.

Within the space of all functions $\phi$  that obey \Eq{diagEQ}, the correct solution is the one that has the lowest value of ${\cal H}$. The latter {\em is not} equal to $\sum_{an}\epsilon^a_n$, since this will count the Coulomb interaction twice. Instead we find
\begin{eqnarray}
{\cal H}_{\rm solution}/N &=& \frac1{2M}\sum_{n=1}^{B+L_s/2}\,\sum_{a=1}^M\left(\epsilon^n_a +  \sum_x \left[{\rm Im}\,  \left(\phi^n_a(x) \phi^{n\star}_a(x+1)\right) + m\, (-1)^x\ \phi^n_a(x) \phi^{n\star}_a(x)\right] \right)\nonumber\\
&+&\frac{g^2N(B+L_s/2)}{4}  \frac1{L_sM^2} \sum'_{abl} \frac1{4\sin^2\left(\frac{2\pi}{L_s}(\frac{a-b}{M}+l)/2\right)}.\label{Hsol}
\end{eqnarray}

\section{Comparison to other relevant works}
\label{otherworks}

In this section we wish to discuss the way that our resulting equations differ from those appearing in other works that also regularize the `t Hooft model on a  finite circle $L$.

\begin{enumerate}
\item Ref.~\cite{LTYL} : this work looked at QCD$_{1+1}$ with the light-cone Hamiltonian, but neglected the curvilinear character of the variables $\varphi$. Thus the  ground state with $\varphi_a=0$ was chosen, that in fact has zero measure within the correct solution. This is equivalent to assuming the Polyakov loops are all nonzero, and in Section~\ref{decompact} we discuss the consequences of such a choice, but in essence it violates volume reduction. In fact, Ref.~\cite{LTYL} showed that this erroneous ground state leads to a phase transition as a function of $L$, which clearly contradicts large-$N$ volume independence.
\item Ref.~\cite{LNT} looked at the Hamiltonian of QCD in axial gauge (in any number of dimensions), and in the continuum. It pointed out to the error made in \cite{LTYL} but did not discuss the consequences of this on the $1+1$ solution presented in \cite{LTYL}. In our paper here we generalize the theoretical framework developed in \cite{LNT} to the lattice regularization. 
\item Ref.~\cite{SchonThies_decompact} : This interesting paper formed one of the motivations for our study. Here the authors showed how the phase transition found in \cite{LTYL} disappears when one chooses an appropriate ansatz for the ground state. The way this choice is made in Ref.~\cite{SchonThies_decompact}, however, is not manifestly gauge invariant and breaks a residual gauge symmetry of the Hamiltonian. This has the disadvantage of making it hard to argue that the ansatz used is exact at large-$N$, and to construct a gauge invariant fermionic ground state.
We avoid this issue in our paper by working with manifestly gauge invariant operators and states. This is most naturally  done with the coherent state approach. Other differences between our paper and \cite{SchonThies_decompact} is the fact that we use the lattice regularization, we do not restrict to zero baryon number or to a particular ansatz for the ground state, and we analyze the role of translation symmetry for volume reduction. In particular, in the following section, we show how a `soft' form of large-$N$ volume independence works if translation invariance breaks to one of its subgroups. Finally we make the connection with the Eguchi-Kawai volume independence manifest. 

\item Ref.~\cite{Salcedo} : To our knowledge this paper is the only one that solves the `t Hooft equation at nonzero baryon number for a general ratio $\sqrt{g^2N}/m$.\footnote{Another paper that discusses nonzero baryon number in the `t Hooft model is \cite{SchonThies}, but there the authors restrict to translation invariant dense systems, which as they show, is inconsistent. For translation non-invariant states, the authors discuss only the vicinity of the chiral limit. As we already mentioned, the fact that the baryons are massless in this limit makes them very different from the four-dimensional QCD case, and it is important to study the $m\neq 0$ case.} Unfortunately, the authors restrict to study a single baryon only, and, like Ref.~\cite{LTYL}, ignore the curvilinear nature of the $\varphi$ coordinates. Consequently, at short lengths (or large baryon densities) their approach would fail, and exhibit the same phase transition seen in \cite{LTYL}. Clearly this calls for revisiting of the topic which we aim to perform in Ref.~\cite{nonzeroBpaper}. 
\end{enumerate}

The list we give above demonstrates the usefulness of our current paper : Firstly, it uses the coherent state approach, which is manifestly gauge invariant throughout. Secondly, it generalize all former studies to the lattice regularization, and extends the study of volume independence to systems with nonzero baryon density and partial translation invariance. Thirdly, it opens the way to study the `t Hooft model for arbitrary values of the quark mass, spatial volume, and Baryon number. To our knowledge this is the first time these steps are taken.

\section{Large-$N$ volume independence}
\label{decompact}

It is generally expected of large-$N$ QCD in $1+1$ dimensions to be independent of its volume. This equivalence of large-$N$ gauge theories with different volumes was first suggested in \cite{EK} and caused great excitement since it was seen to be a potentially easy way to solve large-$N$ QCD on the lattice. Shortly after \cite{EK}, the papers \cite{BHN1,MK} showed that this equivalence breaks down in the continuum limit  for three or more space-time dimensions. This breaking of reduction is signaled by the fact that Polyakov loops that wrap around the volume acquire nonzero expectation values. 

It is useful to put this large-$N$ equivalence in the more general context of orbifold projections between mother and daughter gauge theories : in our case the mother theory is large volume QCD while the daughter theory is small volume QCD. These projections are expected to become equivalences when the rank of the gauge group, $N$, becomes large. For a review on this topic we refer to Ref.~\cite{AEK}. As shown there, these equivalences hold only between certain sectors of the mother and daughter theories which are defined to be neutral under certain symmetries. In the original Eguchi-Kawai paper it was stressed explicitly that the ground state in both theories needs to be symmetric under the center of the gauge group. For $SU(N)$ gauge theories this means that the global $Z_N$ subgroups of the local $SU(N)$, that correspond to multiplying Polyakov loops in different directions by a $Z_N$ phase $e^{2\pi i/N}$, must be unbroken. Thus these Polyakov loops are the order parameters of these symmetries and must have vanishing expectation values for the equivalence to hold.

Another symmetry that the ground states of the mother and daughter theories should respect in the volume projection case is translation symmetry. This is clear intuitively -- how can we describe a theory which breaks translations and that as a result has operators with expectation values that depends on the space-time coordinate, by a theory that has no volume ? The requirement of intact translation symmetry can of course be anticipated from the construction of Ref.~\cite{AEK}, since this is one of the symmetries that define that neutral sectors of a large-to-small volume mapping. Indeed this was already explicitly pointed out in the first paper in \cite{YaffeCoherent}. When this symmetry breaks it is no longer true that the physical observables in the large volume theory have a one-to-one mapping to observables in the zero volume theory. If the attempt to map the large and small volume theories fails, then clearly they cannot be large-$N$ equivalent.
Unfortunately, we believe that the role of translation symmetry in the Eguchi-Kawai equivalence is not fully stressed in some of the relevant literature. The reason, however, is obvious : the QCD vacuum respects translation symmetry ! 

But there is at least one physical scenario where one can expect to get broken translation symmetry in QCD and that is at nonzero baryon number or chemical potential, where crystals of different sort can form (for relevant literature on the topic we refer the interested reader to the review in \cite{MP}). What happens to large-$N$ volume independence in that case ? To answer this question we find it useful to see what happens in a well defined and systematic calculation, and the choice of this paper is the `t Hooft model.\footnote{We note that the following confusion may arise : a tool one can use to make measurements in small volume theories is the Gross-Kitazawa `momentum feeding' trick presented in \cite{GK}. For example, this was used successfully in \cite{KNN}. This trick allows one to extract, from a zero volume theory, the meson propagator $G(x)$ for any value of the separation $x$, and to measure the meson mass from the exponential decay in $|x|$. This reflects how large-$N$ projections repackage (but not lose) the large-volume degrees of freedom into the color indices. A natural question now appears : can one also `repackage', in a similar way, a baryon crystal that breaks translations into the color degrees of freedom of a zero volume theory ? As we shall see below the answer to this is no. This confusion arises because $G(x)$, that depends on $x$, {\em can} be calculated from zero-volume. This, however, is a direct result of translation invariance. When the latter is broken, the meson propagator depends on two space-time coordinates, and the Gross-Kitazawa trick cannot be used.}

Thus our goal in this section is to show how the volume dependent `t Hooft classical Hamiltonian ${\cal H}$, derived in \Eq{Hclassical}, behaves in different cases. We first study the case of a translational invariant ground state (for any value of $B$). We do so for both the original Hamiltonian and also ask what happens if one forces the Polyakov loops to acquire expectation values of different sorts. We then move to discuss what happens when we allow translation symmetry to break.

\subsection{Full translation symmetry}
\label{symm}

Since the staggered fermion Hamiltonian is invariant to translations by two lattice sites, it will be easier to  discuss the original `naive fermions' case in these sections. These fermions are two-component spinors and it is straight forward to repeat the analysis in Section~\ref{SectorF} for them. The result is that $\rho^Q_{xy}$ becomes $2\times 2$ matrix and that the classical Hamiltonian in \Eq{HKC2} now has the following form (here and below the trace refers to this extra $2$-dimensional Dirac space). 
\begin{eqnarray}
{\cal H}_F/N&=&-\frac{i}2 \sum_x \, \tr \, \sigma_3 \, \rho^{0}_{x,x+1} + c.c. + m \sum_x \tr \sigma_1 \, \rho^{0}_{x,x}\nonumber\\
&+&\frac{g^2N}8\, \sum_{x,y\in Z_{L_s}\atop Q\in Z_N}\,\tr \rho^Q_{xy}\, \rho^{-Q}_{yx} \, |x-y+QL_s| e^{-\epsilon|x-y+QL_s|}.\label{HKC5}
\end{eqnarray}
Note that to get \Eq{HKC5} we dropped the constant term that appears when one uses \Eq{rhobar} to write ${\cal H}_F$ in terms of only $\rho$ (and not $\bar\rho$).

Restricting to translation invariant states mean that the coherent states have
\begin{equation}
{\cal C}^{xy}_F = {\cal C}^{x-y}_F.
\end{equation}
Using Eqs.~(\ref{rho}) and~(\ref{thetaBC}) this means that $\rho^Q_{xy}$ is a function of the combination $x-y+QL_s$ :
\begin{equation}
\rho^Q_{xy} \equiv \rho(x-y+QL_s).
\end{equation}
Combining the sums over $x$ and $Q$ into a single sum over the integers, we get
\begin{eqnarray}
{\cal H}_F/(NL_s)&=&\tr \left[ \left(-\frac{i}2  \, \rho(-1)\, \sigma_3 + h.c. + m \, \sigma_1\,\rho(0)\right) \right]+\frac{g^2N}8 \sum_{r\in Z}\tr  \rho(r)\, \rho(r)\, |r|e^{-\epsilon|r|}.
\label{HH}
\end{eqnarray}
Since $\rho(r)$ is defined for all integer values $r$, we can write
\begin{equation}
\rho(r) = \int \frac{dp}{2\pi} e^{ipr } \rho(p).
\end{equation}
In terms of $\rho(p)$ the constraint \Eq{r31}, applied to naive fermions, becomes
\begin{equation}
\int \frac{dp}{2\pi} \, \tr \rho(p) = n_B + 1,\label{r311}
\end{equation}
with $n_B$ equal to the baryon density $B/L_s$, 
and the classical Hamiltonian has the form
\begin{eqnarray}
{\cal H}_F/(NL_s) &=& \int_0^{2\pi} \frac{dp}{2\pi}\, \tr \left[\rho(p) \left( -\sigma_3 \sin(p) + m\, \sigma_1 \right) \right] -\frac{g^2 N}{4}  \pv \pv \frac{dp}{2\pi} \, \frac{dq}{2\pi}\, \frac{\tr \left(\rho(p)\, \rho(q)\right)}{4\sin^2\left((p-q)/2\right)}.\label{HFrp}
\end{eqnarray} 
Here by $\pv$ we mean the principle value which now has a precise meaning in the form of the primed sum in \Eq{HKC4}.

\bigskip 

 The crucial point that we want to make is that Eqs.~(\ref{r311})--(\ref{HFrp}) are independent of $L_s$ and this is the way volume reduction works in our model.\footnote{We note in passing that since all physical information on the system is encoded in the classical Hamiltonian, then the theory's excitation spectrum will also be independent of the volume. For example, the meson spectrum, which is encoded in the $1/N$ fluctuations around the minimum of ${\cal H}$, will also be independent of the volume. For further details on how to extract the spectrum of mesons and glueballs from ${\cal H}$, we refer to Ref.~\cite{YaffeCoherent}.}

\bigskip

It is easy to repeat the above steps for a gluonic state that gives nonzero expectation value to some windings of the Polyakov loop. This can be realized by adding a potential for the Polyakov loops, in similar lines to the potential suggested in \cite{DEK} (although, of course, with an opposite sign since the potential of \cite{DEK} was devised to null all expectation values of all Polyakov loops). A potential can be chosen such that it induces, in the gluon sector, a spontaneous breaking of the $Z_N$ symmetry of the form
\begin{equation}
Z_N\longrightarrow \O.
\end{equation}
In this case we have 
\begin{equation}
\<P_q\>_G = 1, \quad \forall q,
\end{equation}
and it can be easily shown that the only change this causes to \Eq{HKC5} is to replace $\rho^Q_{xy}$ by the $Q$-independent function $\tilde \rho_{xy}$, that is given by 
\begin{equation}
\tilde \rho_{xy} = \sum_{Q\in Z} \rho^Q_{xy}.\label{tilderho}
\end{equation}
Next, for a translational invariant state we write 
\begin{equation}
\tilde \rho_{xy} = \frac1{L_s} \sum_{l=1}^{L_s} \tilde \rho(l)\, e^{i2\pi l/L_s  (x-y)}.
\end{equation}
Since $\tilde \rho$ is now independent of $Q$, one can use \Eq{FT} to perform the sum over $Q$ in \Eq{HKC5} and one gets
\begin{eqnarray}
{\cal H}_F/(NL_s) &=& \frac1{L_s} \sum_{l=1}^{L_s} \tr \left[\tilde \rho(l) ( -\sigma_3 \sin\left(\frac{2\pi l}{L_s}\right) + m\, \sigma_1 ) \right] -\frac{g^2 N}{4}  \frac1{L^2_s}\sum_{l,k=1\atop l\neq k}^{L_s} \, \frac{\tr \left(\rho(l)\, \rho(k)\right)}{4\sin^2\left(\pi (l-k)/L_s\right)}.\nonumber \\ \label{HFrp1}
\end{eqnarray}
The important point about \Eq{HFrp1} is that it depends on $L_s$ is a very strong way.
Indeed, this is the reason Ref.~\cite{LTYL}, which sets all Polyakov loops to be nonzero, saw a strong volume dependence of physical observables which was realized in a phase transition that occurs as a  function of $L$. It is also easy to generalize this result to a breaking of $Z_N\to Z_K$. In that case $L_s$ in the l.h.s. of \Eq{HFrp1} is replaced by $KL_s$, which again means that ${\cal H}_F/L_s$  depends strongly on $L_s$.

An important remark noted also in Ref.~\cite{SchonThies_decompact} here is that when we resolved the quantum Gauss law we assumed that the following conditions holds 
\begin{equation}
\varphi_a - \varphi_b \neq 2\pi n\quad ;\quad n\in Z\label{restrict}
\end{equation}
This is certainly correct for the correct solution with vanishing Polyakov loops where the field configurations that do not obey  \Eq{restrict} have zero Jacobian and thus zero measure (see \Eq{VDM}). For nonzero Polyakov loops, however, \Eq{restrict} can indeed be violated. This means that more zero modes (except for $E^{''I}$) will be present. Since this is not the main topic of this paper we do not treat these here.

\bigskip 

\begin{figure}[hbt]
\centerline{
\includegraphics[width=10cm]{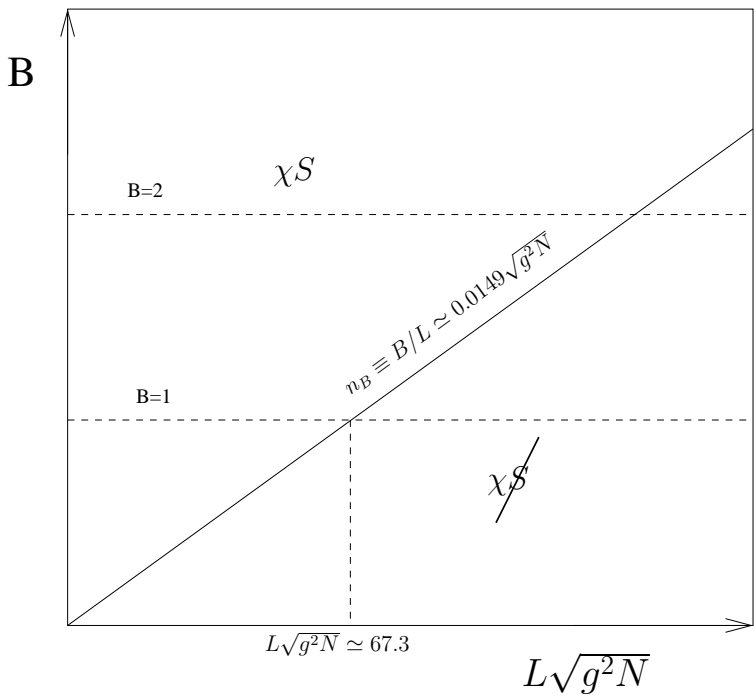}
}
\caption{Phase diagram for chiral symmetry restoration {\em if one assumes translation invariance}. Along fixed lines of fixed $n_B$ there is exact volume independence. See \cite{SchonThies} for more details on the numeric value of the slope of the phase transition line.
}
\label{phaseBL}
\end{figure}

To conclude, if we assume translation symmetry and unbroken $Z_N$, then two systems with the same $n_B$, yet different volumes, will be large-$N$ equivalent.
Ref.~\cite{SchonThies} showed, in the continuum regularization, that in the translation invariant sector, there is a density $n^c_B$ above which chiral symmetry is restored. Let us denote the dimensionless combination $n^c_B/\sqrt{g^2N}$ by $x^c$.\footnote{Ref.~\cite{SchonThies} get $x^c\simeq 0.0149$.} This can be visualized in a simple phase diagram in the space $B$---$L\sqrt{g^2N}$ that we present in Fig.~(\ref{phaseBL}).
Physics along lines of fixed slope is $L$ independent. Nonetheless along horizontal lines, that have a fixed baryon number, one encounters strong $L$ dependence realized as a phase transition. The only case where these two lines coincide is, of course, the $B=0$ case.
We note in passing that the authors of \cite{SchonThies} have shown that the translational invariant ansatz is inconsistent and suggest that the true ground state must break translations. This fits well with the calculation of Ref.~\cite{Salcedo} for the $B=1$ case, and we find preliminary results for $B>1$ which are consistent with this \cite{nonzeroBpaper}.

\subsection{Broken translation symmetry and a crystal of baryons}
\label{asymm}

In this section we wish to emphasize that the volume independence obtained above crucially depends on the assumption that 
\begin{equation}
\rho^Q_{xy} = \rho_{x-y+QL_s},\label{translations}
\end{equation}
i.e. that the ground state is translation invariant. This condition is anticipated in advance from the point of view of orbifold projections \cite{AEK} : the symmetry by which one projects QCD at large volume to QCD at small volume is translation invariance. 

\begin{figure}[bt]
\centerline{
\includegraphics[width=10cm]{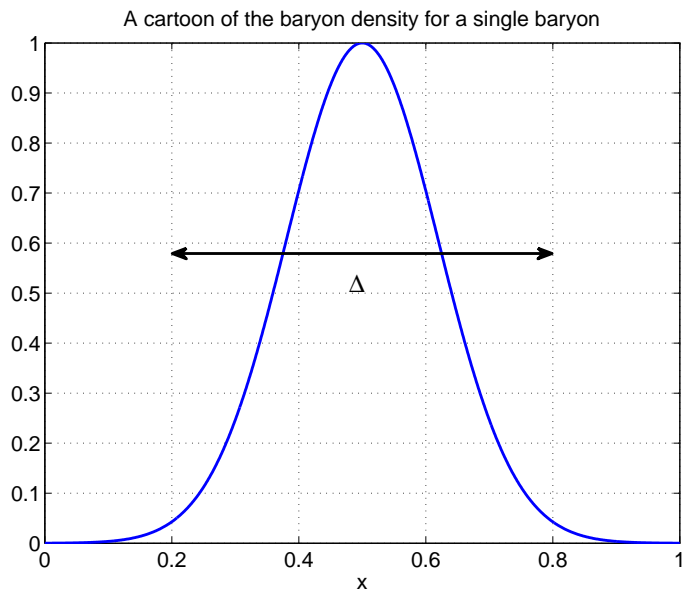}
}
\caption{A cartoon of the baryon density of a single baryon at large-$N$ where a complete breakdown of translation invariance takes place and large-$N$ volume reduction does not work at all.
}
\label{B1}
\end{figure}

\begin{figure}[bt]
\centerline{
\includegraphics[width=10cm]{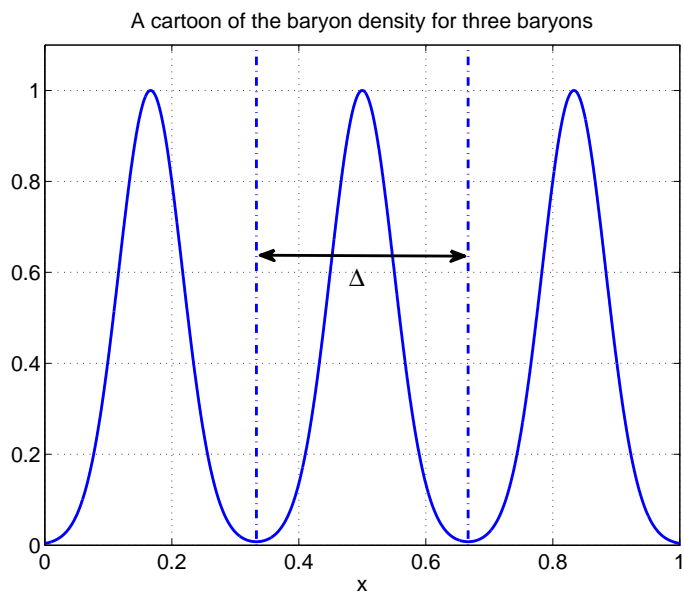}
}
\caption{A cartoon of the baryon density of three baryons at large-$N$ -- a partial breakdown of translation invariance : translations by a third of a unit still leave this state invariant. Thus large-$N$ reduction holds only between this system and a system with a single baryon on a box of size $\Delta$.
}
\label{Bcrystal}
\end{figure}

It is useful to make a concrete example to demonstrate this point, and we first consider the case of single baryon in a box. At large-$N$ the baryon has mass of $O(N)$ and is a static particle\footnote{We do not consider the chiral limit here, where two dimensional baryons become massless.}. As a result, its presence in the system spontaneously breaks translation invariance. Indeed the calculation in Ref.~\cite{Salcedo} showed that the baryon wave function in the `t Hooft model pictorially looks like Fig.~\ref{B1}, where we present a sketch of the baryon density. 
In Ref.~\cite{nonzeroBpaper} we plan to revisit this calculation and to extend it to a finite baryon density. Let us assume that for sufficiently large $B$ a crystal of baryons will form - of the sort sketched in Fig.~\ref{Bcrystal}. 
In both the single baryon and the baryon crystal cases the ground state wave function has a characteristic length scale $\Delta$ -- it is the baryon width for $B=1$ and the baryon-crystal wave-length for $B\gg 1$. These length scales must depend on $g^2N$, $m$, and $L$, and it is clear that decreasing the box size $L$ will change the ground state wave functions -- the baryons will get squashed. This is a result of the compactness of space and the periodic boundary conditions. The only case where the ground state represented by Figs.~(\ref{B1}-\ref{Bcrystal}) is invariant under a change in $L$ is if $\Delta=\infty$ which corresponds to intact translational symmetry. 

Nonetheless, there is an interesting subtlety to the statement above : in contrast to the single baryon case, the baryon crystal sketched in Fig.~\ref{Bcrystal} has an unbroken symmetry  : shifts by $\Delta$ leave the ground state invariant.  Thus the following softer form of large-$N$ volume-independence survives in this case. 

\bigskip
\bigskip
{\em Two systems $I$ and $II$ defined in volumes $L_I$ and $L_{II}$, and accommodating baryon crystals with baryon numbers $B_I$ and $B_{II}$, are large-$N$ equivalent if their baryon densities $n_{I,II}=B_{I,II}/L_{I,II}$ are equal.}
\bigskip
\bigskip

The meaning of this statement is depicted pictorially in Fig.~\ref{Bcrystal} : the three baryon system forms a crystal in a box of unit size, and so in these units $\Delta=1/3$. The statement above means we can reproduce the physics of this three-baryon system from a single-baryon system whose box size is $1/3$ -- the system bounded between the two dash-dot vertical lines of Fig.~\ref{Bcrystal}. In these units the baryon number density of both systems is $3$, and this is the only relevant parameter at large-$N$.
Let us now show how this happens. 

\bigskip

The starting point is to modify the ansatz in \Eq{translations} for the way $\rho^Q_{xy}$ depends on $Q,x$, and $y$. It will be first useful to change coordinates and define 
\begin{eqnarray}
r &\equiv & x + QL -y , \\
s &\equiv& x+ QL +y,\\
\rho_{rs} &\equiv&\rho^Q_{xy}.
\end{eqnarray}
Since $x,y \in [1,L_s]$ and $Q\in Z$, we have $r \in Z$. Given one fixes the value of $r$, the values $s$ obtains are $r+2,r+4,r+6,\dots,r+2L_s$. The periodicity in $\Delta$ means that
\begin{eqnarray}
\rho_{r,s+\Delta} &=& \rho_{r,s}.
\end{eqnarray}
One solution to this condition is (here we assume that both $L_s/\Delta$ and $L_s$ are even)
\begin{equation}
\rho_{r,s} = \int \frac{dp}{2\pi} \sum_{q\in Z_{2\Delta}} \, \rho_q(p)\, e^{ipr + iq\frac{2\pi}{2\Delta}s}.\label{FTnew}
\end{equation}
Substituting \Eq{FTnew} into the first two terms of \Eq{HKC5} we have
\begin{eqnarray}
1^{\rm st}-{\rm term} &=& \int\, \frac{dp}{2\pi}\, \tr \left[ -\sigma_3 \sin(p) \left( \rho_{2\Delta}(p) - \rho_{\Delta}(p)\right)\right],\label{H1st}\\
 2^{\rm nd}-{\rm term} &=&  \int\, \frac{dp}{2\pi}\, \tr \left[ m \sigma_1\, \left( \rho_{2\Delta}(p) + \rho_{\Delta}(p)\right)\right].\label{H2nd}
\end{eqnarray}
Substituting \Eq{FTnew} into the third term of ${\cal H}_F$ gives
\begin{eqnarray}
3^{\rm rd}-{\rm term}/(NL_s)&=& -\frac{g^2N}{4}\sum_{q=1}^{2\Delta}\, \sum_{k=1}^{4}\pv \pv \frac{dp}{2\pi}\frac{dp'}{2\pi}\, \left[ \frac{\tr (\rho_q(p) \, \rho_{k\Delta-q}(p')}{4\sin^2((p-p')/2 + \pi k/2)} \right],
\label{H3rd}
\end{eqnarray}
Where we define $\rho_{q<0}(p)=\rho_{q>2\Delta}(p)=0$, and $\rho_{0}(p)\equiv\rho_{2\Delta}(p)$. The sum over $k$ is a result of the summation over $x=(r+s)/2$ and $y=(s-r)/2$ that one performs upon the substitution of \Eq{FTnew} into the Coulomb interaction term in the Hamiltonian. Finally the global Gauss Law constraint applied on $\rho_{rs}$ becomes
\begin{equation}
\int \frac{dp}{2\pi} \, \tr \left(\rho_\Delta(p) + \rho_{2\Delta}(p)\right) = n_B+1.
\end{equation}
As it stands, the classical `t Hooft Hamiltonian density ${\cal H}_F/(NL_s)$ and the Gauss law constraint, depend only on the parameters $m$, $g^2N$, $n_B$, and $\Delta$ -- the volume is irrelevant at large-$N$.

Similarly to what we saw in Section~\ref{symm}, this volume independence is a direct result of the ansatz we took in \Eq{FTnew} that reflects a partial translation symmetry, and of the fact that all nontrivial winding of the spatial Polyakov loops have a zero expectation value. If we were to neglect the zero modes then all Polyakov loops would effectively be equal and nonzero. This will cause the infinite set of density matrices $\rho^Q_{xy}$ to be replaced by the $Q$-independent quantity $\tilde \rho_{xy}$ (see \Eq{tilderho}) which depends only on the two compact coordinates $x$ and $y$. As a result the coordinate $r=x-y$ would have been as well, and \Eq{FTnew} will be replaced by 
\begin{equation}
\rho_{r,s} = \frac{1}{L_s}\sum_{l=1}^{L_s} \,\sum_{q\in Z_{2\Delta}}\, \rho_{q,l}\, e^{2\pi \,i\, (lr/L_s + qs/2\Delta)}. 
\end{equation}
Plugging this into the classical Hamiltonian will result in a function that strongly depends on $L_s$ and this is wrong.

\subsection{Remark on lattice simulations of volume-reduced large-$N$ QCD at nonzero $B$ or $\mu$}
\label{remark}

Finally, let us make a remark of relevance to lattice practitioners. Consider large-$N$ QCD defined in a very small box with side $L$ that obeys $L\ll 1/\Lambda_{QCD}$. Let us also assume reduction holds for zero baryon number (this assumption is automatically fulfilled  in two dimensions and for higher dimensions is expected to hold in modified versions of QCD -- see \cite{DEK}). Now, force the system to accommodate a single or a few baryons. This can be done by working in the canonical ensemble, or by increasing the chemical potential in the grand canonical ensemble. Because the volume is small, the baryon density will be huge, $n_B\gg (\Lambda_{QCD})^d$ (here $d$ is the number of spatial dimensions), and what we prove in the previous section tells us that at large-$N$ this small volume system is equivalent to a `standard' large-volume system which is extremely dense. For a lattice theory the density will be one in units of the cutoff and the following consequence is immediate :
trying to study baryons with single-site reduced models of the Eguchi-Kawai type drives the theory towards the saturation regime of the lattice, where the density is of $O(1/a)$. This regime is dominated by lattice artifacts and is unphysical. 

This may well be (partly) the reason reason why Ref.~\cite{GHN}, which works with a single site model at nonzero chemical potential $\mu$, sees either a ground state that is empty of baryons (for small $\mu$) or a ground state that is full of baryons, with density of $O(1/a)$, and  that disappears in the continuum limit. As we discuss above, such behavior is expected from a single-site construction, and cannot be used to study the physical regime where the baryon density is not at the cutoff scale.

\section{Conclusions, some remarks, and an outlook}
\label{summary}

In this paper we study the way large-$N$ QCD depends on its volume. General arguments, such as those found in Ref.~\cite{AEK} and in its references, tell us what are the requirements that the ground state of a large-$N$  gauge theory needs in order to be volume independent. Nonetheless, we find it is useful to see how this phenomenon emerges explicitly in a theory which is exactly soluble. This is the reason we chose to study the `t Hooft model in this paper.

The formalism we use is the lattice Hamiltonian formalism in axial gauge. Since we are working with a finite box size, the gauge fields cannot be gauged away completely and, in the gluon sector, one is left with a set of $N-1$ curvilinear quantum zero modes. These describe the $N-1$ eigenvalues of the spatial Polyakov loop. Together with their conjugate momenta these zero modes determine the leading $O(N^2)$ dynamics. Thus a systematic large-$N$ treatment of the full Hamiltonian proceeds as follows:
\begin{enumerate}
\item Treat the pure gauge Hamiltonian -- find the so called `large-$N$ master field'.
\item Solve for the dynamics of the fermions. They now interact on the background of this master field.
\end{enumerate} 
Thus, the back-reaction of the fermions on the gauge fields, which is subleading in $O(1/N)$ compared to the gauge fields, is neglected and this is consistent as long as these $1/N$ effects are not enhanced by any massless modes, which we assume to be the case.

To perform steps (1-2) above we choose to use the lattice UV regularization and so generalized the axial gauge fixing of Ref.~\cite{LNT} in the Hamiltonian to the lattice. 
We solve the fermion sector with the coherent state approach of Ref.~\cite{YaffeCoherent}. The reason we choose this approach is that it is manifestly gauge invariant and easy to justify at large-$N$. We then end up with a regularized form of the `t Hooft Hamiltonian that explicitly depends on traces of the Polyakov loops, and describes their interaction with quark-antiquark pairs in the background of baryons.

Our next step was to analyze the volume dependence of the emerging `t Hooft Hamiltonian. We showed that if translation symmetry is intact then
\begin{enumerate}
\item When the $Z_N$ symmetry, whose order parameters are the spatial Polyakov loops, is intact, then the spatial coordinate in the `t Hooft Hamiltonian decompactifies and volume independence emerges. 
\item In contrast, when the $Z_N$ symmetry breaks to $Z_K$, then the `t Hooft Hamiltonian has a strong volume dependence. We emphasize again that this analysis ignores a set of additional zero modes that appear in this case. Since this is not the main topic of this paper we do not study this issue further, but the reader should be aware of this point. (The focus of this paper was the case with unbroken center symmetry, which is free from this subtlety)
\end{enumerate}
In our case, the gauge dynamics tell us that the Polyakov loops vanish, the $Z_N$ is intact,  and so volume reduction takes place.

A important component in the validity of volume reduction is the fact that the ground state is translation invariant. In our calculation we see how this condition arises explicitly. Moreover in the case that translation invariance breaks down by a crystal of baryons, we show that a softer form of volume independence takes place and that, at large-$N$, instead of studying a crystal of $B$ baryons in a volume $L$, one can study a single baryon in a volume $L/B$. In both cases the baryon number density is the same and together with the gauge coupling and the quark mass, these are the only relevant parameters determining the properties of the large-$N$ ground state -- the volume is irrelevant.

Another aspect of large-$N$ gauge theories which is explicitly exposed in this work is the dominance of the gauge fields dynamics over the fermions dynamics at large-$N$, and that it also happens at nonzero baryon number. This means that the physics of the ground state is planar and that quark loops are suppressed. This is in contrast to the conjecture raised in Ref.~\cite{Cohen}, where the author suggests that quark loops are important at nonzero baryon chemical potential, even at $N=\infty$. This conjecture was originally proposed to resolve a subtle apparent confusing contradiction between standard diagrammatic large-$N$ arguments and the phenomenology of QCD. This confusion is absent from our approach to the two-dimensional case, and we see that the conjecture of Ref.~\cite{Cohen} does not hold there. Briefly, the way this contradiction gets resolved is by non-perturbative effects, and so using perturbation theory (even if it is planar) is quite misleading. This  means that the `contradiction' of large-$N$ and phenomenology is only apparent also in four dimensions. Further discussion on this point will be given in Ref.~\cite{nonzeroMUpaper}.

\bigskip
\bigskip
To conclude, we show how  volume independence emerges in a clear and simple
 way in the Hamiltonian approach to the `t Hooft model in the lattice regularization. In the presence of a baryon crystal a partial form of volume independence  allows one to substitute the study of the crystal of wave-length  $\Delta$, baryon number $B$, and box of size $L$, with a system of a single baryon in a box of size $L=\Delta$. The latter may be useful in our companion study~\cite{nonzeroBpaper} where we aim to solve the `t Hooft Hamiltonian of \Eq{Hclassical} given a baryon number $B\ge 1$. Surprisingly, this has not been done yet for arbitrary quark mass (for the vicinity of the chiral limit where the two-dimensional baryons are nearly massless, see \cite{SchonThies}). In fact, even in the $B=1$ case, which was studied in Ref.~\cite{Salcedo}, the spatial Polyakov loops were set to unity. As discussed above this is inconsistent with the gauge sector dynamics and will give erroneous results at small enough volume (and by the equivalence mention above, at large enough densities, if one increases the density by fixing the baryon number and decreasing the volume). In \cite{nonzeroBpaper} we also plan to see what will be the effect of correcting this issue. 

Finally we hope that knowing how two-dimensional QCD behaves at nonzero baryon number will be of value for studies of the physical four-dimensional system. In particular it seems that the partial independence of the QCD ground state on the volume, that we see emerging in the $1+1$ case, is of a general nature and from the orbifold projection point of view can be anticipated on general grounds. This makes the successful modification to the Eguchi-Kawai reduction the four-dimensional case (of the type of \cite{DEK}) appealing, since it will allow one to study a single baryon in a modestly-sized box (but not of zero size) and conclude on how large-$N$ four-dimensional QCD behaves at moderate/high densities.

\section*{Acknowledgments}
I thank R.~Narayanan for correspondence during several stages of this work and for discussing his related work \cite{GHN}. I am grateful to L.~Yaffe for numerous enlightening discussions on the coherent state approach and on large-$N$ volume independence. I also thank S.~R.~Sharpe for comments on this draft and for discussions on staggered fermions, to Carlos Hoyos-Badajoz for discussions on matrix models and related issues, and to V.~P.~Nair for an interesting discussion on IR divergences. This work was supported in part by the U.S. Department of Energy under Grant No. DE-FG02-96ER40956.

\appendix

\section{Hermiticity of ${\cal E}_a$ and its commutation relations}
\label{gauge_fixing}

In this Appendix we show that the zero mode $E^{''I}_{p=0}$ obeys
\begin{eqnarray}
\left(E^{''I}_{p=0}\right)^{\dag}&=&E^{''I}_{p=0},\label{zeroherm}\\
\left[E^{''I}_{p=0},E^{''J}_{p=0}\right] &=&0\label{zerocomm}.
\end{eqnarray}
To show \Eq{zeroherm} we use \Eq{E''def} and write the difference between $E^{''I}_{p=0}$ and its hermitian conjugate
\begin{eqnarray}
\left(E^{''I}_{p=0}\right)^{\dag}-E^{''I}_{p=0} &=& \frac1{\sqrt{L_s}} \sum_x\, \left[ E^{'i}_x , \left(V^{\rm Adj.}_x\right)_{iI} \right] = \frac1{\sqrt{L_s}} \sum_x \left[ E^{'i}_x,\left(U^\dag_{x-1}\cdots U^\dag_1 S^\dag e^{i\varphi x}\right)_{iI}\right]\nonumber\\
&=&\frac1{\sqrt{L_s}} \sum_x \left(U^\dag_{x-1}\cdots U^\dag_1\right)_{ij}\left[E^{'i}_x,\left(S^\dag e^{i\varphi x}\right)_{jI} \right]\nonumber \\
&=&\frac1{\sqrt{L_s}} \sum_x \left(U^\dag_{x-1}\cdots U^\dag_1\right)_{ij}\left(V^{\rm Adj.}_x\right)_{ik}\left[E^{''k}_x,\left(S^\dag e^{i\varphi x}\right)_{jI} \right]\nonumber \\
&=&\frac1{\sqrt{L_s}} \sum_x \left(U^\dag_{x-1}\cdots U^\dag_1\right)_{ij}\left(V^{\rm Adj.}_x\right)_{iK}S^\dag_{jl}\left[E^{''K}_{p=0},\left(e^{i\varphi x}\right)_{lI} \right].
\end{eqnarray}
To get the last line we used the fact that, within the physical Hilbert space, and except for $E^{''k}_{p=0}$ with $k\in [1,N-1]$, all other components of the operator $E^{''}$ commute with the gauge fields (see Eqs.~(\ref{E''p}) and~(\ref{Glaw''1})), and that $E^{''I}$ commute with the $S$ operators. Finally, it is easy to show that 
\begin{equation}
\left(e^{i\varphi x}\right)_{lI}=\delta_{lI},
\end{equation}
which proves that $E^{''I}_{p=0}$ is hermitian within the physical sector of Hilbert space. Since $\lambda^I_{aa}$ is real this also means that ${\cal E}_a$ is hermitian.

Let us now show that \Eq{zerocomm} is obeyed within this subspace. We begin by writing the l.h.s in terms of $E^{'}$. The result is :
\begin{eqnarray}
\frac1{L_s}\sum_{xy}\left[E^{''I}_x,E^{''J}_y\right] &=& \frac1{L_s}\sum_{xy}\left(V^{\rm Adj.}_x\right)_{iI} \left(V^{\rm Adj.}_y\right)_{jJ} \left[E^{'i}_x,E^{'j}_y \right] \nonumber \\
&+&\frac1{L_s}\sum_{xy}\left\{\left(V^{\rm Adj.}_x\right)_{iI} \left[E^{'i}_x,\left(V^{\rm Adj.}_y\right)_{jJ} \right] E^{'j}_y - 
\left(
I\leftrightarrow J
\right)
\right\}\nonumber \\
\label{rhs}
\end{eqnarray}
Using the commutation relations of $E'$ (which are the same as those of $E$ -- see \Eq{E'}) it is easy to show that the first term on the r.h.s. is given by
\begin{equation}
1^{\rm st}{\rm - term} = \frac1{L_s} \sum_x \sum_k \left(V^{\dag}_x T^k V_x\right)^{\rm Adj.}_{JI} E^{'k}_x.
\end{equation}
Here the c-number matrix $T_k$ is the $k^{\rm th}$ generator of $SU(N)$ in the adjoint representation. Also, because $I,J \in [1,N-1]$, we can write this term as
\begin{eqnarray}
1^{\rm st}{\rm - term} &=& \frac1{L_s} \sum_x \sum_k \left(\Omega^k_x \right)_{JI} E^{'k}_x,\\
\left(\Omega^k_x\right)_{IJ}&\equiv& \left(SU_{1}\cdots U_{x-1}\, T^k \,U^\dag_{x-1}\cdots U^\dag_1 S^\dag\right)_{JI}.\label{Omega}
\end{eqnarray}

Next we proceed to the second term and first evaluate the commutation relation of $E^{'i}_x$ with $\left(V^{\rm Adj.}_x\right)_{jJ}$ : because $J\in [1,N-1]$ then $\left(V_y\right)_{jJ}=\left(U^\dag_{y-1} \cdots U^\dag_1 S^\dag \right)_{jJ}$, and we have 
\begin{eqnarray}
\left[E^{'i}_x,\left(V^{\rm Adj.}_y\right)_{jJ}\right]&=&\left(U^\dag_{y-1} \cdots U^\dag_1 \right)_{jl} \left[E^{'i}_x,S^\dag_{lJ}\right] + {\cal C}_F(y-1-x) \left[E^{'i}_x,\left(U_1\cdots U_{y-1}\right)_{lj}\right]S^\dag_{jJ}\nonumber\\
&=&\left(U^\dag_{y-1} \cdots U^\dag_1 \right)_{jl} \left(V^\dag_x\right)_{qi} \left[E^{''q}_x,S^\dag_{lJ}\right] \nonumber\\
&+&{\cal C}_F(y-1-x) \left( U_1 \cdots U_{x-1}\right)_{lf}\left[E^{'i}_x,\left(U_{x}\right)_{fg}\right]\left(U_{x+1}\cdots U_{y-1}\right)_{gj}S^\dag_{jJ}.\label{commE'V}
\end{eqnarray}
The first term vanishes since all $E^{''}$ fields commute with the $S$ fields. We now need the commutation relation between $E^{'}$ and $\left(U_x\right)^{\rm Adj.}$. We calculated them explicitly by using \Eq{comm_plus1}, and after some algebra we find an expected result :
\begin{equation}
\left[E^{'i}_x,\left(U^{\rm Adj}_x\right)_{fg}\right] = \left(T^i U^{\rm Adj.}_x\right)_{fg}.
\end{equation}
Using this we obtain the following expression for the $2^{\rm nd}$ term of the r.h.s. of \Eq{rhs}
\begin{equation}
2^{\rm nd}{\rm -term} = \frac1{L_s}\sum_{x<y}\left(V_x\right)_{iI} \left(SU_1 \cdots U_{x-1} T^i U_x U_{x+1} \cdots U_{y-1}\right)_{Jj}E^{'j}_y - (I\leftrightarrow J).\label{2nd1}
\end{equation}
Since $T^{i}_{lk}=-T^{k}_{li}$ we can rearrange \Eq{2nd1} to get
\begin{equation}
2^{\rm nd}{\rm -term} = -\frac1{L_s}\sum_{x<y} \left[\left(\Omega^k_x \right)_{JI}-\left(\Omega^k_x \right)_{IJ}\right] \left(U_{x}\cdots U_{y-1}\right)_{kj}E^{'j}_y = -\frac2{L_s}\sum_{x<y} \left(\Omega^k_x\right)_{JI}\left(U_x \cdots U_{y-1}\right)_{kj}\, E^{'j}_{y}.\label{2nd2}
\end{equation}
Here we got the last equality by using the fact that $\Omega_{IJ} = -\left(\Omega\right)_{JI}$ (which is a result of $T^k_{ij}=-T^{j}_{ji}$).

The outcome of the above paragraphs is that the commutation relations between $E^{''I}_{p=0}$ and $E^{''J}_{p=0}$ are proportional to a linear combination of $\left(\Omega^k_x\right)_{IJ}$.
Let us now show that $(\Omega^k_x)_{ij}=0$ if $i,j\in [1,N-1]$. For that  we basically write the definition of $\Omega$ : 
\begin{equation}
\left(\Omega^k_x\right)_{ij}=\tr \left[\lambda^i \, S U_1 \cdots U_{x-1} \lambda^g U^\dag \cdots U^\dag_1 S^{\dag} \right] \, T^{k}_{gf} \,\tr \left[\lambda^f \, U^\dag_{x-1}\cdots U^\dag_1 S^\dag \, \lambda^j S U_1 \cdots U_{x-1} \right], 
\end{equation}
where here the trace is in the fundamental representation, and we use the fact that $T^k_{gf}=\frac12 \tr \lambda^f\left[\lambda^g,\lambda^k\right]$ and the completeness relation of the $\lambda$ matrices. After some algebra we get 
\begin{equation}
\left(\Omega^k_x\right)_{ij}=\frac12 \tr \left(\left[\lambda^i , \lambda^j\right] S U_1 \cdots U_{x-1} \lambda^k U^\dag_{x-1} \cdots U^\dag_1 S^\dag\right),
\end{equation}
which shows that for $i,j$ belonging to the {\em abelian} Cartan Sub-algebra, then $\left(\Omega^k\right)_{ij}=0$, and that consequently
\begin{equation}
\left[E^{''I}_{p=0},E^{''J}_{p=0}\right]=0.
\end{equation}
Clearly this also means that $[{\cal E}_a,{\cal E}_b]=0$.

\section{Calculation of ${\cal H}_F$.}
\label{appHF}
In this section we present the calculation of ${\cal H}_F({\cal C}_F,{\cal P})$. We begin by evaluating the expectation value of the numerator $\rho_{ab}(x)\rho_{ba}(y)$ of \Eq{HC} in the fermionic coherent state $|{\cal C}_F\>={\cal U}({\cal C}_F)|0\>$, where the reference state $|0\>$ is defined in Section~\ref{SectorF}. For that we write
\begin{equation}
\<{\cal C}_F|\rho_{ab}(x)\rho_{ba}(y)|{\cal C}_F\> = \<0|\,{\cal U}^\dag({\cal C}_F)\, \rho_{ab}(x)\, {\cal U}({\cal C}_F)\, {\cal U}^\dag({\cal C}_F)\, \rho_{ba}(y) \,{\cal U}({\cal C}_F)\,|0\>.
\end{equation}
To calculate ${\cal U}^\dag({\cal C}_F)\, \rho_{ab}(x)\, {\cal U}({\cal C}_F)$ we write
\begin{eqnarray}
{\cal U}^\dag({\cal C}_F)\, \rho_{ab}(x)\, {\cal U}({\cal C}_F) &=& \frac12 \, \left[ {\cal U}^\dag({\cal C}_F)\, \psi^{\dag b}_x \, {\cal U}({\cal C}_F)\,   {\cal U}^\dag({\cal C}_F)\, \psi^{a}_x \, {\cal U}({\cal C}_F) \right. \nonumber \\
&& \left. -\frac{\delta_{ab}}{N}\sum_c \, {\cal U}^\dag({\cal C}_F)\, \psi^{\dag c}_x \, {\cal U}({\cal C}_F)\,   {\cal U}^\dag({\cal C}_F)\, \psi^{c}_x \, {\cal U}({\cal C}_F)\right],
\end{eqnarray}
and use the  Hadamard lemma to show that
\begin{equation}
{\cal U}^\dag ({\cal C}_F) \, \psi^a_x \, \,{\cal U}({\cal C}_F) = \sum_{y\in Z} \, e^{i\varphi_a (y-x)/L_s}\, \left(e^{-i{\cal C}_F}\right)^{xy} \, \psi^a_y.\label{UpsiU}
\end{equation}
This then gives
\begin{eqnarray}
\<{\cal C}_F|\rho_{ab}(x)\rho_{ba}(y)|{\cal C}_F\> &=& \frac14 \sum_{vw\in Z\atop v'w'\in Z} \left(e^{-i{\cal C}_F}\right)^{xv}\left(e^{i{\cal C}_F}\right)^{wx}\left(e^{-i{\cal C}_F}\right)^{yv'}\left(e^{i{\cal C}_F}\right)^{w'y}\nonumber\\
&&\times \<0|\left[ e^{i\varphi_a(v-x)/L_s + i\varphi_b(x-w)/L_s}\, \psi^{\dag b}_w\psi^a_v - \frac{\delta_{ab}}N\sum_c\,e^{i\varphi_c(v-w)/L_s}\psi^{c\dag}_w\psi^c_v\right]\nonumber\\
&&\times \left[ e^{i\varphi_b(v-y)/L_s + i\varphi_a(y-w')/L_s}\, \psi^{\dag b}_{w'}\psi^a_{v'} - \frac{\delta_{ab}}N\sum_c\,e^{i\varphi_c(v'-w')/L_s}\psi^{c\dag}_{w'}\psi^c_{v'}\right]|0\>.\nonumber \\ \label{rhorho}
\end{eqnarray}
We proceed we need to evaluate fermionic contractions of three types:
\begin{itemize}
\item Terms of type I : for the $a\neq b$ terms of \Eq{HC} we need to evaluate $\<0|\psi^{\dag b}_w\, \psi^a_v\,\psi^{\dag a}_{w'}\psi^b_{v'}|0\>$. Since $\psi^a_{\tilde z}|0\>=\psi^{\dag a}_z|0\>=0$ for $z\in S$ and $\tilde z\in \bar S$  we get
\begin{equation}
\<0|\psi^{\dag b}_w\, \psi^a_v\,\psi^{\dag a}_{w'}\psi^b_{v'}|0\> = \sum_{z\in S\atop \tilde z\in \bar S}\delta_{\bar w z} \delta_{\bar v \tilde z} \delta_{\bar{w}'\tilde z}\delta_{\bar{v}'z}.\label{contract1}
\end{equation}
Here we use the notation where $\bar{x}=x \,{\rm mod}\,L_s$ (recall that $S$($\bar S$) are the set of sites that are full(empty) of quarks).
\item Terms of type II : for the $a= b$ terms of \Eq{HC} we need to evaluate $\<0|\psi^{\dag a}_w\, \psi^a_v\,\psi^{\dag a}_{w'}\psi^a_{v'}|0\>$. Here we have more contractions and we get
\begin{equation}
\<0|\psi^{\dag a}_w\, \psi^a_v\,\psi^{\dag a}_{w'}\psi^a_{v'}|0\> = \sum_{z\in S\atop \tilde z\in \bar S}\delta_{\bar w z} \delta_{\bar v \tilde z} \delta_{\bar{w}'\tilde z}\delta_{\bar{v}'z} + \sum_{z_1\in S\atop z_2\in S}\delta_{\bar w z_2} \delta_{\bar v  z_2} \delta_{\bar{w'}z_1}\delta_{\bar{v}'z_1}.\label{contract2}
\end{equation}
\item Terms of type III : for the $a= b$ terms of \Eq{HC} we also need to evaluate $\<0|\psi^{\dag a}_w\, \psi^a_v\,\psi^{\dag c}_{w'}\psi^c_{v'}|0\>$, with $a\neq c$. Here we have only one  contractions that gives
\begin{equation}
\<0|\psi^{\dag a}_w\, \psi^a_v\,\psi^{\dag c}_{w'}\psi^c_{v'}|0\> = \sum_{z_1\in S\atop z_2\in S}\delta_{\bar w z_2} \delta_{\bar v  z_2} \delta_{\bar{w'}z_1}\delta_{\bar{v}'z_1}.\label{contract3}
\end{equation}
\end{itemize}

To express \Eq{rhorho} in a compact form we define the following `matrix densities'
\begin{eqnarray}
\rho^q_{xy} &=& \sum_{p\in Z\atop z\in S} \left(e^{i{\cal C}_F}\right)^{z+pL_s,x} \, \left(e^{-i{\cal C}_F}\right)^{y,z+(p+q)L_s},\\
\bar \rho^{q}_{xy} &=& \sum_{p\in  Z\atop \tilde z\in \bar S} \left(e^{i{\cal C}_F}\right)^{\tilde z+pL_s,x} \, \left(e^{-i{\cal C}_F}\right)^{y,\tilde z+(p+q)L_s}.
\end{eqnarray}
The hermiticity and periodicity of the matrix ${\cal C}_F$ imply that these matrix densities obey 
\begin{eqnarray}
\bar \rho^q_{xy} &=& \delta_{xy}\delta_{q,0} - \rho^q_{xy},\label{r11}\\
\sum_{q\in Z\atop y\in L_s} \rho^q_{xy} \rho^{-q}_{yz} &=& \rho^{q=0}_{xz},\label{r21}\\
\sum_{x\in L_s} \rho^{q}_{xx} &=& (B+L_s/2)\delta_{q,0}.\label{r31}
\end{eqnarray}

Substituting Eqs.~(\ref{contract1})--(\ref{contract3}) into \Eq{rhorho} we find  that $\<{\cal C}_F|H_C|{\cal C}_F\>$ can be brought to the form
\begin{eqnarray}
\<{\cal C}_F|H_C|{\cal C}_F\> &=&\frac{g^2}{4L_s} \sum_{l=1}^{L_s} \sum_{a\neq b} \sum_{qq'} \frac{\rho^{q'}_{xy}\, \bar \rho^{-q}_{yx} \, e^{-i2\pi l (x-y)/L }}{4\sin^2((\varphi_a-\varphi_b)/L_s + 2\pi l/L_s)/2} \times  e^{-i\varphi_a(x-y +q L_s)/L_s + i\varphi_b( x-y + q'L_s)/L_s} \nonumber \\
&+&\frac{g^2}{4L_s}\sum_{l=1}^{L_s-1} \sum_{qq'} \frac{(1-\frac1{N} ) e^{i2\pi l/L (y-x)}}{4\sin^2(\pi l/L_s)} \left[ \rho^q_{xy}\bar \rho^{-q'}_{yx} + \rho^q_{xx} \rho^{-q'}_{yy}\right]\times \sum_a e^{i\varphi (q-q')},\nonumber \\
&-&\frac{g^2}{4L_sN}\sum_{l=1}^{L_s-1}\sum_{qq'}\frac{\rho^q_{xx}\rho^{-q'}_{yy}e^{i2\pi l/L_s(y-x)}}{4\sin^2(\pi l/L_s)} \times \sum_{a\neq c} e^{i\varphi_a q -i\varphi_c q'}. \label{HC1}
\end{eqnarray}

To proceed we first show that the terms in the last two lines of \Eq{HC1} are subleading. To see this note that we still need to perform the integral of \Eq{Hclassical}. This will be done conveniently if we write these terms as 
\begin{eqnarray}
&&\frac{g^2N}{4L_s}\left\{\sum_{l=1}^{L_s-1} \sum_{qq'} \frac{(1-\frac1{N} ) e^{i2\pi l/L (y-x)}}{4\sin^2(\pi l/L_s)} \left[ \rho^q_{xy}\bar \rho^{-q'}_{yx} + \rho^q_{xx} \rho^{-q'}_{yy}\right]\times P^{q-q'}\right.
\nonumber \\
&-&\left.\sum_{l=1}^{L_s-1}\sum_{qq'}\frac{\rho^q_{xx}\rho^{-q'}_{yy}e^{i2\pi l/L_s(y-x)}}{4\sin^2(\pi l/L_s)} \times \left(P_q \, P_{-q'} - \frac1{N} P_{q-q'}\right)\right\}. \label{2terms}
\end{eqnarray}
Here we have defined the $k$-wound Polyakov loop operator as
\begin{equation}
P_k\equiv \frac1{N} \sum_a e^{i\varphi_a k}.
\end{equation}

The averages over the Haar measure of $P_k$ and $P_k \times P_{k'}$ are known explicitly for the  $U(N)$  group \cite{diaconis,rains}
\begin{eqnarray}
\<P_k\> &=& \delta_{k,0},\label{Pk}\\
\<P_k\, P_{-k'}\> &=& \delta_{kk'}\, \left(\delta_{k,0} + \frac{\min(|k|,N)}{N^2}\right).\label{PkPk'}
\end{eqnarray}
Using these results we can see that both these terms are at most of $O(1)$. 
For an $SU(N)$ group these averages are expected to differ by a small amount since the center of mass of the eigenvalues $\sum_c\varphi_c$ is held fixed at zero. For example, the average of $P_k$ can be done explicitly : one can expand the Vandermond as a polynomial of $e^{i\varphi_a}$ and see that $\<P_k\>_{SU(N)}=0$ if $|k| \ge 2N$. Together with the $Z_N$ symmetry of the Vandermond measure and the methods in chapter 8 of \cite{creutzbook} we get
\begin{equation}
\<P_k\> = \delta_{k,0} + \frac1{N} \delta_{|k|,N},\label{Pksun}
\end{equation}

In contrast to the two subleading terms we discussed above, the first term in \Eq{2terms} includes a double sum over color indices and is thus of $O(g^2N^2)\sim N$. The same is true for the scaling of kinetic contribution of the fermions to the classical Hamiltonian : $\<{\cal C}_F|H_K|{\cal C}_F\>$. Using \Eq{UpsiU} it is easy to show that it is given by
\begin{equation}
\<{\cal C}_F|H_K|{\cal C}_F\> =N \sum_x \left[\left(-\frac{i}2 \rho^q_{x,x+1}  + c.c. \right)+ m (-1)^x \rho^q_{xx} \right] \times P_q.
\end{equation}
which simplifies even further if we use \Eq{Pksun}.

\section{Resolving the constraints on $\rho^q_{xy}$}
\label{app_rho_resolve}

In this appendix we wish to show how the constraints in Eqs.~(\ref{r11}--\ref{r31}) are resolved. 
Our starting point is to write the Fourier transform  
\begin{equation}
\rho^q_{xy} = \int_{-\pi}^\pi \, \frac{dp}{2\pi}\, \rho_{xy}(p)\, e^{ip\, q}.
\end{equation}
This is the most general way to express the dependence of $\rho^q_{xy}$ on the integer $q$ since the  range of the latter is the whole integers. In terms of $\rho_{xy}(p)$ the constraints of Eqs.~(\ref{r11}--\ref{r31}) become the following infinite set of equations that hold for any value of $p\in (0,2\pi]$.
\begin{eqnarray}
\rho_{xy}(p) &=& \rho^\star_{yx}(p),\label{r11p}\\
\sum_{x=1}^{L_s}\, \rho_{xx}(p) &=& \left(B + L_s/2\right),\label{r21p}\\
\sum_y \rho_{xy}(p)\, \rho_{yz}(p)&=&\rho_{xz}(p).\label{r31p}
\end{eqnarray}
The equations above can be solved by setting up, for each $p$,  an orthogonal basis on the $L_s\times L_s$ space that is furnished by the indices $x$ and $y$. We denote the orthogonal wave functions on that basis by $\phi^n_x(p)$, with $n=1,2,\dots,L_s$. In terms of these wave functions the hermitian matrices $\left(\rho(p)\right)_{xy}$ can be written as
\begin{equation}
\rho_{xy}(p) = \sum_{n=1}^{L_s}\, \phi^n_x(p)\, r_n(p)\, \phi^{n\star}_x(p), 
\end{equation}
where $r_n(p)$ are the eigenvalues of the matrix $\rho(p)$ in the basis spanned by $\phi(p)$. Using \Eq{r11p} we see that the eigenvalues $r_n(p)$ are real, while using \Eq{r31p}, we see that they obeys
\begin{equation}
r_n(p)\left( r_n(p)-1\right)=0.
\end{equation}
Thus $r_n(p)$ is either $0$ or $1$. Finally, \Eq{r21p} tells us that 
\begin{equation}
\sum_{n=1}^{L_s} \, r_n(p) = B + L_s/2, \quad ; \quad \forall p,
\end{equation}
and by ordering the eigenfunctions $\phi^n(p)$ according to their values of $r_n(p)$ we see that 
\begin{equation}
r_n(p) = \left[ 
\begin{array}{lr}
1 & \qquad n \le B+L_s/2,\\
0 & \qquad n > B+L_s/2.
\end{array}
\right.
\end{equation}
The end product of the discussion above is that the most general way to resolve Eqs.~(\ref{r11}--\ref{r31}) is to write
\begin{equation}
\rho^q_{xy} = \int_{0}^{2\pi}\, \frac{dp}{2\pi}\, \sum_{n=1}^{B+L_s/2}\, \phi^n_x(p) \, \phi^{n\star}_y(p)\, e^{ipq},\label{rho_resolve_1}
\end{equation}
with the functions $\phi^n_x(p)$ forming an orthogonal basis for each value of $p$ (note that this basis is not necessarily the same for each $p$ and the precise form of $\phi^n_x(p)$ is determined by the dynamics of the ${\cal H}$ minimization).

\bigskip 

The last step is only done for the convenience of numerically minimizing  ${\cal H}$ with respect to $\rho$  : we discretize the momentum space $p\in (0,2\pi]$ by using the standard form
\begin{equation}
p \quad \stackrel{{\rm discretize}}{\longrightarrow} \quad \frac{2\pi a}{M}\, \quad {\rm with} \quad a=1,2,\dots,M.
\end{equation}
This discretization, together with \Eq{rho_resolve_1} gives Eqs.~(\ref{constraints_res1}--\ref{constraints_res2}) and becomes equivalent to the exact system only in the limit of $M\to \infty$.


\begin{thebibliography}{99}
\bibitem{largeN} G.~'t Hooft, 
  Nucl.\ Phys.\  B {\bf 72}, 461 (1974).
  Nucl.\ Phys.\ B {\bf 75}, 461 (1974);  
E.~Witten, 
Nucl.\ Phys.\ B {\bf 160}, 57 (1979);  
  A.~V.~Manohar, 
arXiv:hep-ph/9802419.  
For a recent review see K.~Peeters and M.~Zamaklar, 
arXiv:0708.1502 [hep-ph]. 

\bibitem{lattice-reviews} M.~Teper, 
[arXiv:hep-lat/0509019];
R.~Narayanan and H.~Neuberger, 
arXiv:0710.0098 [hep-lat].  

\bibitem{EK} T.~Eguchi and H.~Kawai, 
Phys.\ Rev.\ Lett.\ {\bf 48}, 1063 (1982). 

\bibitem{BHN1} G.~Bhanot, U.~M.~Heller and H.~Neuberger, 
Phys.\ Lett.\ B {\bf 113}, 47 (1982). 

\bibitem{MK}
  V.~A.~Kazakov and A.~A.~Migdal,
  Phys.\ Lett.\  B {\bf 116}, 423 (1982).

\bibitem{Migdal}
  A.~A.~Migdal,
  Phys.\ Lett.\  B {\bf 116}, 425 (1982).

\bibitem{TEK} A.~Gonzalez-Arroyo and M.~Okawa, 
Phys.\ Lett.\ B {\bf 120}, 174 (1983).

\bibitem{AEK} P.~Kovtun, M.~Unsal and L.~G.~Yaffe, 
JHEP {\bf  0706}, 019 (2007) [arXiv:hep-th/0702021]. 

\bibitem{DW} S.~R.~Das and S.~R.~Wadia, 
Phys.\ Lett.\  B {\bf 117}, 228 (1982) [Erratum-ibid.\ B {\bf 121}, 456 (1983)].

\bibitem{GK} D.~J.~Gross and Y.~Kitazawa, 
Nucl.\ Phys.\ B {\bf 206}, 440 (1982).  

\bibitem{Parisipapers}
 G.~Parisi,  
Phys.\ Lett.\ B {\bf 112}, 463 (1982).  
G.~Parisi and Y.~C.~Zhang, 
Nucl.\ Phys.\ B {\bf 216}, 408 (1983).  
G.~Parisi and Y.~C.~Zhang, 
Lett.\ B {\bf 114} (1982) 319.  


\bibitem{KNN} J.~Kiskis, R.~Narayanan and H.~Neuberger, 
Phys.\ Rev.\ D {\bf 66}, 025019 (2002) [arXiv:hep-lat/0203005].


\bibitem{QEK} 
B.~Bringoltz and S.~R.~Sharpe,
  arXiv:0810.1239 [hep-lat].
Phys.\ Rev.\ D {\bf 78}, 034507 (2008)  [arXiv:0805.2146 [hep-lat]]

\bibitem{DEK}
 M.~Unsal and L.~G.~Yaffe, 
  Phys.\ Rev.\  D {\bf 78}, 065035 (2008)
  [arXiv:0803.0344 [hep-th]].


\bibitem{SchonThies_decompact}
  V.~Schon and M.~Thies,
  Phys.\ Lett.\  B {\bf 481}, 299 (2000)
  [arXiv:hep-th/0001162].

\bibitem{nonzeroBpaper}
  B.~Bringoltz,
  arXiv:0901.4035 [hep-lat].


\bibitem{Cohen}
  T.~D.~Cohen,
  Phys.\ Rev.\  D {\bf 70}, 116009 (2004)
  [arXiv:hep-ph/0410156].

\bibitem{Salcedo}
  L.~L.~Salcedo, S.~Levit and J.~W.~Negele,
  Nucl.\ Phys.\  B {\bf 361}, 585 (1991).

\bibitem{SchonThies}
  V.~Schon and M.~Thies,
  Phys.\ Rev.\  D {\bf 62}, 096002 (2000)
  [arXiv:hep-th/0003195].
  arXiv:hep-th/0008175.

\bibitem{nonzeroMUpaper}
B.~Bringoltz, 
work in progress.



\bibitem{plethora}
I.~Bars and M.~B.~Green,
  Phys.\ Rev.\  D {\bf 17}, 537 (1978).
  M.~Li,
  Phys.\ Rev.\  D {\bf 34}, 3888 (1986).
  M.~Engelhardt,
  Phys.\ Lett.\  B {\bf 355}, 507 (1995)
  [arXiv:hep-th/9503155].
  A.~R.~Zhitnitsky,
  Phys.\ Lett.\  B {\bf 165}, 405 (1985)
  [Sov.\ J.\ Nucl.\ Phys.\  {\bf 43}, 999.1986\ YAFIA,43,1553 (1986\ YAFIA,43,1553-1563.1986)].
  L.~D.~McLerran and A.~Sen,
  Phys.\ Rev.\  D {\bf 32}, 2794 (1985).
  F.~Lenz, M.~A.~Shifman and M.~Thies,
  Phys.\ Rev.\  D {\bf 51}, 7060 (1995)
  [arXiv:hep-th/9412113].
  M.~Engelhardt,
  Phys.\ Lett.\  B {\bf 355}, 507 (1995)
  [arXiv:hep-th/9503155].



\bibitem{LTYL}
  F.~Lenz, M.~Thies, K.~Yazaki and S.~Levit,
  Annals Phys.\  {\bf 208}, 1 (1991).
\bibitem{LNT}
  F.~Lenz, H.~W.~L.~Naus and M.~Thies,
  Annals Phys.\  {\bf 233} (1994) 317.

\bibitem{MEandBEN}
  B.~Bringoltz and B.~Svetitsky,
  Phys.\ Rev.\  D {\bf 68}, 034501 (2003)
  [arXiv:hep-lat/0211018].

\bibitem{YaffeCoherent}
  L.~G.~Yaffe,
  Rev.\ Mod.\ Phys.\  {\bf 54}, 407 (1982).
  F.~R.~Brown and L.~G.~Yaffe,
  Nucl.\ Phys.\  B {\bf 271}, 267 (1986).
  T.~A.~Dickens, U.~J.~Lindqwister, W.~R.~Somsky and L.~G.~Yaffe,
  Nucl.\ Phys.\  B {\bf 309}, 1 (1988).





\bibitem{Kogut75}
  J.~B.~Kogut and L.~Susskind,
  Phys.\ Rev.\  D {\bf 11}, 395 (1975).

\bibitem{Susskind}
L.~Susskind,
  Phys.\ Rev.\  D {\bf 16}, 3031 (1977).

\bibitem{Douglas}
  M.~R.~Douglas,
  Nucl.\ Phys.\ Proc.\ Suppl.\  {\bf 41}, 66 (1995)
  [arXiv:hep-th/9409098].





\bibitem{MP}
  L.~McLerran and R.~D.~Pisarski,
  Nucl.\ Phys.\  A {\bf 796}, 83 (2007)
  [arXiv:0706.2191 [hep-ph]].





\bibitem{diaconis}
P.~Diaconis and S.~E.~Evans,
Trans.\ Amer.\ Math. \ Soc. {\bf 353}, 2615-2633 (2001).
\bibitem{rains}
E.~M.~Rains, 
Probab.\ Theory Related Fields\ {\bf 107} (1997), 219-241
\bibitem{creutzbook}
  M.~Creutz,
{\it  Cambridge, Uk: Univ. Pr. ( 1983) 169 P. ( Cambridge Monographs On Mathematical Physics)}

\bibitem{GHN}
  R.~Galvez, A.~Hietanen and R.~Narayanan,
  Phys.\ Lett.\  B {\bf 672}, 376 (2009)
  [arXiv:0812.3449 [hep-lat]].

\end{thebibliography}
\end{document}